\documentclass[12pt]{article}
\usepackage{amscd}
\usepackage{bbm}
\usepackage{mathrsfs}
\usepackage{amssymb}
\usepackage{graphics}
\usepackage{graphicx}
\usepackage{amsfonts}
\usepackage{amsmath}
\usepackage{caption}
\usepackage{titlesec}
\usepackage{subcaption}
\usepackage{float}
\usepackage[numbers,sort&compress]{natbib}
\titleformat{\section}{\large\bfseries}{\thesection. }{0em}{}
\titleformat{\subsection}{\normalsize\bfseries}{\thesubsection. }{0em}{}
\numberwithin{equation}{section}
\begin{document}
	\date{}
	\title{Breather interactions and limit analysis in the second harmonic generation process via Riemann-Hilbert approach}
	
	\author{An-Yao Jin, Rui Guo$ \thanks{Corresponding author,
	    	gr81@sina.com}$ \
		\\
		\\{\em
			School of Mathematics, Taiyuan University of} \\
		{\em Technology, Taiyuan 030024, China} } \maketitle
	
	\begin{abstract}		
	The discovery of second harmonic generation (SHG) heralds the emergence of nonlinear optics. In this paper, we focus on the theoretical analysis of the SHG equation under phase-matching conditions. A rich family of soliton solutions are derived via the Riemann-Hilbert (RH) approach, and we characterize breather interactions corresponding to second harmonic solutions. The construction and solution of the RH problem are discussed firstly, including a detailed analysis of the discrete spectrum in the single-zero and double-zero cases. In such cases two-soliton solutions, breather solutions, two-breather solutions, and soliton-breather solutions are obtained. We numerically simulate and visually illustrate the spatiotemporal evolution of these solutions. Furthermore, through asymptotic analysis of the interaction dynamics, the exact position shift magnitudes resulting from breather-breather interaction within a nonzero background field are calculated. When the velocities are distinct, the interaction of two breathers divides the $xt$-plane into four asymptotic regions by the characteristic trajectories of breathers, and we show that the asymptotic behavior can be explicitly determined by the relative position between the region and the breathers. 
		
	\vspace{7mm}\noindent\emph{Keywords}: Second harmonic generation equation; Riemann-Hilbert problem; Breather interactions
	\end{abstract}
	\newpage

\section{Introduction}
\hspace{0.7cm}The second harmonic generation (SHG) process \cite{ref1} corresponds to the elementary processes of the annihilation of two light quanta and the creation of one new quantum with twice the energy. The first experimental observation of SHG was achieved by irradiating a single crystal of quartz with intense coherent light from a pulsed ruby optical maser \cite{ref2}. Usually the SHG process can be governed by coupled equations \cite{ref3}:
 
\begin{equation}
	\begin{aligned} \label{s11}
	&\frac{dA_1}{dz}=\frac{2i\omega _{1}^{2}d_{\mathrm{eff}}}{k_1c^2}A_2A_{1}^{*}e^{-i\Delta kz},
	\\
	&\frac{dA_2}{dz}=\frac{i\omega _{2}^{2}d_{\mathrm{eff}}}{k_2c^2}A_{1}^{2}e^{i\Delta kz},
	\end{aligned}
\end{equation}\\
 where $A_1$ and $A_2$ are the field amplitudes of the fundamental frequency (FF) and second harmonic (SH), respectively. $\omega_1$, $k_1$ are the frequency and propagation constant for FF, and $\omega_2$, $k_2$ are the corresponding quantities for the SH. $\Delta k = 2k_1-k_2$ is the wavevector mismatch, $d_{\mathrm{eff}}$ represents nonlinear susceptibility, $z$ is propagation distance, and $c$ is the speed of light in vacuum. Eq.~(\ref{s11}) describes the fundamental mechanism by which a non-centrosymmetric crystal, when subjected to an optical field, generates a nonlinear polarization that produces radiation at the second-harmonic frequency.
 
 Generally, the signal of SHG is extremely weak and practically negligible, rendering it of little practical value. Until later, Giordmaine observed a sizable intensity of the second harmonic signal in potassium dihydrogen phosphate without usage of the focused beams \cite{ref4}. This discovery demonstrated that phase matching enables waves to achieve fully coherent superposition along the propagation path, thereby promoting the widespread application of SHG in numerous fields such as laser technology \cite{ref5,ref6}, materials research \cite{ref7,ref8,ref9}, biological imaging \cite{ref10}, and nonlinear optics \cite{ref11,ref12}. In particular, compared to traditional methods, the application of the SHG technology has completely broken the reliance on a vacuum environment in material analysis and has gradually become an irreplaceable tool. 

Based on the phase-matching conditions, the coupled equation (\ref{s11}) can be transformed through straightforward algebra, which ultimately yielded the following relatively simple and integrable form \cite{ref13,ref14}:
\begin{align} \label{s12}
	\partial _{\chi}q_1=-2q_2q_{1}^{*}, \ \partial _{\tau}q_2=q_{1}^{2}.
\end{align}
The solution of Eq.~(\ref{s12}) under specific boundary conditions has long been a subject of significant interest \cite{ref15,ref16,ref17,ref18,ref19,ref20}. Since the Cauchy and Goursat problems for the SHG equation were defined in Ref.~\cite{ref17}, the solution of the Goursat problem over different intervals has been continuously explored through inverse scattering transform (IST) \cite{ref18,ref19}. On finite intervals the Goursat problem was solved by introducing an effective S-matrix \cite{ref18}, and the Weyl function was employed to yield explicit solutions on the semi-strip \cite{ref19}. The lack of linear dispersion inherent to Eq.~(\ref{s12}) makes constructing solutions on the infinite interval highly challenging. Building upon prior work, an approximate solution was obtained by solving the Gel'fand-Levitan-Marchenko (GLM) equation, and the asymptotic limit of the solution for large $\tau$ was analyzed \cite{ref20}. Differing from this approach, we will define boundary conditions distinct from previous ones and solve the SHG equation on the entire complex plane by constructing the corresponding Riemann-Hilbert (RH) problem. The RH method has been widely applied to the study of integrable systems since it was first introduced \cite{ref21,ref22,ref23,ref24,ref25,ref26}. This approach can yield novel soliton and breather solutions distinct from those previously obtained and also shows the graphical representation of the solutions. 

Using the RH approach, a rich family of soliton solutions for the coupled system has been computed \cite{ref27,ref28,ref29,ref30,ref31}. The interactions between solitons were first studied by the nonlinear Schrödinger equation in the defocused medium \cite{ref32}, and this work was further extended to the coupled system under non-zero boundary conditions \cite{ref33,ref34,ref35,ref36}. By solving the second harmonic solutions $q_2$ in the single-pole case, plenty of breather solutions can be obtained. We will analyze the characteristics of breather interactions based on different $xt$-plane regions. Specifically, we will conduct a further analysis from five cases depending on the value of velocities.

The main work of this paper is as follows: In Section 2, we first analyze the analyticity, symmetry, and asymptotic behavior of the Jost eigenfunctions and the scattering data. Based on these properties, we construct and solve the related RH problems and reconstruct the potential functions in both single-pole and double-pole cases. The breather solutions, two-soliton solutions, two-breather solutions, and soliton-breather solutions are presented in the form of images in Section 3, which provide a clearer observation of their dynamical behaviors. In Section 4, differing from the asymptotic analysis of the single-breather solution from three perspectives: $V<V_1,V=V_1,V>V_1$, we also analyze the asymptotic limits of the two-breather solution and calculate the displacement induced by its interaction from five distinct perspectives: $V<V_1,V=V_1,V_1<V<V_2,V=V_2,V>V_2$. Finally, we make a conclusion.

\section{Riemann-Hilbert approach}
\subsection{Direct Scattering}
\hspace{0.7cm}Eq.~(\ref{s12}) admits the Lax pair 
\begin{align}\label{s21}
	V_\chi=XV, \,\, V_\tau=TV,
\end{align}
where
\begin{align}
	X\left( \chi ,\tau,\xi \right) =\left( \begin{matrix}
		-i \xi&		2q_2\\
		2q^*_2&		i\xi\\
	\end{matrix} \right),\,\,  
    T\left( \chi ,\tau ,\xi \right) =\frac{i}{\xi}\left( \begin{matrix}
    	-q_{1}^{*}q_1&		-q_{1}^{2}\\
    	q_{1}^{*2}&		q_{1}^{*}q_1\\
    \end{matrix} \right). \nonumber
\end{align}
\hspace{0.7cm}When $\chi \to \pm \infty$, the asymptotic mixed boundary conditions are given by 
\begin{align}
	\lim_{\chi \to \pm \infty}q_1 = 0, \,\, \lim_{\chi \to \pm \infty}q_2 = \tilde{q}_{2,\pm}  e^{i\alpha}.\nonumber
\end{align}
Replacing $V$ with $\tilde{V}e^{\frac{i\alpha \tau \sigma _3}{2}}$, the asymptotic spectrum problem can be expressed as follows:
\begin{subequations}\label{s22}
	\begin{align}
		&	\tilde{V}_{\pm ,\chi} =X_\pm \tilde{V}_\pm =(-i \xi \sigma_3 + \tilde{X}_\pm)\tilde{V}_\pm,\\
		&	\tilde{V}_{\pm ,\tau} =T_\pm \tilde{V}_\pm={\alpha  2\xi}X_\pm\tilde{V}_\pm,
	\end{align}
\end{subequations}
with
$$
    \sigma _3=\left( \begin{matrix}
    	1&		0\\
    	0&		-1\\
    \end{matrix} \right),\,\,
 	\tilde{X}_\pm=\left( \begin{matrix}
 		0&		2\tilde{q}_{2,\pm}\\
 		2\tilde{q}^*_{2,\pm}&		0\\
 	\end{matrix} \right),
$$ 
 and $\left| \tilde{q}_{2,\pm} \right|=q_{20}$. By analyzing the eigenvalues of $X_\pm$, which satisfy $k^2 = \xi^2 - 4 q^2_{20}$, we find that, whether it is considered in the complex $k$-plane or on the Riemann surface, there is a certain complexity. Consequently, we try to construct RH problem on the complex $z$-plane, where the uniformization variable $z = \xi + k$, and $$k(z)=\frac{1}{2}(z-\frac{4q_{20}^{2}}{z}),\,\, \xi (z)=\frac{1}{2}(z+\frac{4q_{20}^{2}}{z}).  $$  
For further analysis below, the complex $z$-plane is divided into the following three parts:
\begin{subequations}
	\begin{align}
		&D^+ = \{z \in \mathbb{C}: ({\left| z \right|}^2+4q_{20}^2)\text{Im}z >0 \}, \\
		&D^- = \{z \in \mathbb{C}: ({\left| z \right|}^2+4q_{20}^2)\text{Im}z <0 \}, \\
		&\Sigma = \{z \in \mathbb{C}: ({\left| z \right|}^2+4q_{20}^2)\text{Im}z =0 \}.
	\end{align}
\end{subequations}

We can denote asymptotic eigenvector matrix as $Y_{\pm}(z)=I-\frac{i}{z}\sigma _3\tilde{X}_{\pm}$, where $I$ is the identity matrix. As $\chi \rightarrow \pm \infty$, one can define $v_\pm(\chi ,\tau,z) = Y_\pm(z) e^{-i \theta(\chi ,\tau,z)\sigma_3}$, where $\theta (\chi ,\tau ,z)=k(z)(\chi +\frac{\alpha}{2\xi}\tau )$. Here, $v_\pm = \left(v_{\pm,1}, v_{\pm,2}\right)$ with the subscripts 1 and 2 indicating the first and second columns, respectively. Under this asymptotic condition, we introduce $\mu_\pm(\chi ,\tau,z) = v_\pm(\chi ,\tau,z) e^{i \theta(\chi ,\tau,z)\sigma_3}$ so that $\mu_\pm \sim Y_\pm $. Integrating along two special paths, there are Jost eigenfunctions that satisfy
\begin{equation}\label{s23}
	Y _{\pm}^{-1} \mu _{\pm}\left( \chi,\tau,z \right) =I-\int_x^{\pm\infty}{e}^{ik \left(y-x \right) \hat{\sigma}_3}\left[ Y _{\pm}^{-1}\Delta X _{\pm}\left( y,\tau,z \right) \mu _{\pm}\left( y,\tau,z \right) \right] dy,
\end{equation}
with $ \Delta X _{\pm}\left( y,\tau,z \right) = X\left( \chi,\tau,z \right) - X_\pm\left( \chi,\tau,z \right)$.

From Eq.~(\ref{s21}), one can find that $v_{\pm}$ have such a linear relationship as follows:
\begin{align}\label{s24}
	v_+\left( \chi ,\tau ,z \right)=v_-\left( \chi ,\tau ,z \right) S\left( z \right),
\end{align}
where
$$
S\left( z \right) =\left( \begin{matrix}
	s_{11}\left( z \right)&		s_{12}\left( z \right)\\
	s_{21}\left( z \right)&		s_{22}\left( z \right)\\
\end{matrix} \right) .$$
is the scattering matrix, and the scattering matrix satisfies
\begin{subequations}\label{s25}
	\begin{align}		
		s_{11}\left( z \right)=Wr\left( v_{+,1},v_{-,2} \right) /\mathrm{\gamma},\ s_{12}\left( z \right)=Wr\left( v_{+,2},v_{-,2} \right) /\mathrm{\gamma},
		\\ 
		s_{21}\left( z \right)=Wr\left( v_{-,1},v_{+,1} \right) /\mathrm{\gamma},\ s_{22}\left( z \right)=Wr\left( v_{-,1},v_{+,2} \right) /\mathrm{\gamma},
	\end{align}
\end{subequations}
with $\mathrm{\gamma} = Wr(v_{-,1},v_{-,2})$ and $s_{ii}(z)$ as the scattering data, $i=1,2$.

The reflection coefficients are defined by
$$
\rho \left( z \right) =\frac{s_{21}\left( z \right)}{s_{11}\left( z \right)}, \,\,  \tilde{\rho}\left( z \right)=\frac{s_{12}\left( z \right)}{s_{22}\left( z \right)}.
\\
$$

Define $\mu_\pm = \left(\mu_{\pm,1}, \mu_{\pm,2}\right)$ on the real axis, and the two columns of $\mu_{\pm}$ satisfy the following properties, respectively:
	
	\textbf{Proposition 1.} For $q_2-q_{2,\pm}\in L^1\left( \mathbb{R} \right) $, $\mu _{-,1}$, $\mu _{+,2}$ can be analytically extended to $D_{+}$, while $\mu _{-,2}$, $\mu _{+,1}$ can be analytically extended to $D_{-}$. 
	
	 Based on Volterra integral equations (\ref{s23}), constructing Neumann series in the corresponding regions and according to Bellmann inequality, it is easy to show the analyticity of $\mu_{\pm,j}$, $j=1,2$.

	\textbf{Proposition 2.} Considering the following transformations in complex $z$-plane: $(1)$  $z\to z^*$, $(2)$ $z\rightarrow \frac{4q_{20}^{2}}{z}$, the scattering matrix and Jost eigenfunctions satisfy the following symmetries:
	\begin{align}\label{s271}
	&	\mu _{\pm}\left( \chi ,\tau ,z \right) =\sigma _1\mu _{\pm}^{*}\left( \chi ,\tau ,z^* \right) \sigma _1, \mu _{\pm}\left( \chi ,\tau ,z \right) =-\frac{i}{z}\mu _{\pm}\left( \chi ,\tau ,\frac{4q_{20}^{2}}{z} \right) \sigma _3X_{\pm}, \\ \label{s272}		
	&	S\left( z \right) =\sigma_1 S^*\left( z^* \right) \sigma_1,\ S\left( \frac{4 q_{20}^2}{z} \right) ={\sigma_3 X_{-}} S\left( z \right)\left( {\sigma_3 X_{+}}\right)^{-1},
	\end{align}
	where $$\sigma _1=\left( \begin{matrix}			0&		1\\
			1&		0\\
		\end{matrix} \right).$$		
		
\textbf{Proposition 3.} The asymptotic properties of the Jost eigenfunctions and scattering matrix are as follows:
	\begin{align}
		&\mu _{\pm}\left( \chi ,\tau ,z \right) = I-\frac{i}{z}\sigma _3X_{\pm}+O\left( \frac{1}{z} \right),\,\, z\rightarrow \infty,\\
        &\mu _{\pm}\left( \chi ,\tau ,z \right) = -\frac{i}{z}\sigma _3X_{\pm}+O\left( 1 \right),\,\, z\rightarrow 0, \\
        &S\left( z \right) =I+O\left( \frac{1}{z} \right),\,\, z\rightarrow \infty,\\
        &S\left( z \right) =\left( \begin{matrix}
        	\frac{\tilde{q}_{2,-}}{\tilde{q}_{2,+}}&		0\\
        	0&		\frac{\tilde{q}_{2,+}}{\tilde{q}_{2,-}}\\
        \end{matrix} \right),\,\, z\rightarrow 0.
	\end{align}

	\subsection{Riemann-Hilbert problem with single poles}
	\hspace{0.7cm} The symmetries of the reflection coefficients can be derived from the symmetries given by Eq.~(\ref{s272}): 
	\begin{equation}\label{s31}
		\rho \left( z \right) =\tilde{\rho}^*\left( z^* \right) =-\frac{\tilde{q}_{2,-}^{*}}{\tilde{q}_{2,-}}\tilde{\rho}\left( \frac{4q_{20}^{2}}{z} \right) =-\frac{\tilde{q}_{2,-}^{*}}{\tilde{q}_{2,-}}\tilde{\rho}^*\left( \frac{4q_{20}^{2}}{z^*} \right) .
	\end{equation}
    Meantime, it is easy to get
    $$
    s_{11}\left( z_n \right)=s_{11}\left( \frac{4q_{20}^{2}}{z_{n}^{*}} \right) =s_{22}\left( z_{n}^{*} \right)=s_{22}\left( \frac{4q_{20}^{2}}{z_n} \right).
    $$
    Define discrete spectrum $Z=\left\{ \zeta_n,\zeta _{n+N},\zeta _{n}^{*},\zeta _{n+N}^{*} \right\} _{n=1}^{N}$, where $\zeta _n=z_n,\ \zeta _{n+N}=\frac{4q_{20}^{2}}{z_n}.$ Due to the two columns of the $\mu_\pm$ having different analyticity, we construct the following sectionally matrix functions $P^\pm$:
    \begin{align}\label{s32}
    	P^+\left( \chi ,\tau ,z \right) =\left( \mu _{-,1},\frac{\mu _{+,2}}{s_{22}} \right), \, \, P^-\left( \chi ,\tau ,z \right) =\left( \frac{ \mu _{+,1}}{s_{11}},\mu _{-,2}\right),
    \end{align}
    so that they can be extended to the corresponding analytical region, respectively. Expanding the linear relationship given by Eq.~(\ref{s24}), we get
    \begin{align} \label{s33}
    	\frac{\mu _{+,1}}{s_{11}}=\mu _{-,1}+\frac{s_{21}}{s_{11}}\mu _{-,2}e^{2i\theta}, \, \, \mu _{-,2}=\frac{\mu _{+,2}}{s_{22}}-\frac{s_{12}}{s_{22}}\mu _{-,1}e^{-2i\theta}.
    \end{align}
The following RH problem can be derived immediately:\\
(1) $P^\pm$ are meromorphic in $\mathbb{C}\setminus \Sigma$, respectively, \\
(2) $P^-\left( \chi ,\tau ,z \right) =P^+\left( \chi ,\tau ,z \right) \left( I-J(\chi,\tau,z) \right), \, z\in \Sigma $, \\
(3) $P^{\pm}=I+O\left( \frac{1}{z} \right) $, as $z\rightarrow \infty$, $P^{\pm}=-\frac{i}{z}\sigma _3\tilde{X}_-+O\left( 1 \right)$, as $z\rightarrow 0$,\\
where jump matrix is 
$$ J(\chi,\tau,z)=\left( \begin{matrix}
	\rho \left( z \right) \tilde{\rho }\left( z \right)&		\tilde{\rho} \left( z \right)e^{-2i\theta\left( z \right)}\\
	-\rho \left( z \right) e^{2i\theta\left( z \right)}&		0\\
\end{matrix} \right),$$
   
By removing the asymptotic behavior and the pole contributions of $P^\pm$, one has
	\begin{equation}
	\begin{aligned}
	P^- -I+\frac{i}{z}\sigma _3\tilde{X}_-&-\sum_{n=1}^{2N}{\frac{\underset{z=\zeta _{n}^{*}}{\text{Res}}P^+}{z-\zeta _{n}^{*}}}-\sum_{n=1}^{2N}{\frac{\underset{z=\zeta _n}{\text{Res}}P^-}{z-\zeta _n}}\\
	&=P^+ -I+\frac{i}{z}\sigma _3\tilde{X}_--\sum_{n=1}^{2N}{\frac{\underset{z=\zeta _{n}^{*}}{\text{Res}}P^+}{z-\zeta _{n}^{*}}}-\sum_{n=1}^{2N}{\frac{\underset{z=\zeta _n}{\text{Res}}P^-}{z-\zeta _n}}-P^+J,
	\end{aligned}
	\end{equation}
according to Plemelj formula, it is easy to get
\begin{align}\label{s71}
P=I-\frac{i}{z}\sigma _3\tilde{X}_-+\sum_{n=1}^{2N}{\frac{\underset{z=\zeta _{n}^{*}}{\text{Res}}P^+}{z-\zeta _{n}^{*}}}+\sum_{n=1}^{2N}{\frac{\underset{z=\zeta _n}{\text{Res}}P^-}{z-\zeta _n}}+\frac{1}{2\pi i}\int_{\varSigma}{\frac{P^+J}{\zeta -z}}d\zeta .
\end{align}

Eq.~(\ref{s25}) yields
$$
v_{+,1}\left( z_n \right) =b_nv_{-,2}\left( z_n \right),\,\, v_{+,2}\left( z_{n}^{*} \right) =\tilde{b}_nv_{-,1}\left( z_{n}^{*} \right),\,\, n=1,...,2N.
$$
Then, the residue conditions can be expressed as
\begin{subequations}\label{s34}
	\begin{align} 
		&\underset{z=z_n}{\text{Res}}\left[ \frac{\mu _{+,1}\left( \chi ,\tau ,z \right)}{s_{11}\left( z \right)} \right] =C_ne^{2i\theta \left( z_n \right)}\mu _{-,2}\left( \chi ,\tau ,z_n \right),\,\, n=1,...,2N,\\
		&\underset{z= z_{n}^{*}}{\text{Res}}\left[ \frac{\mu _{+,2}\left( \chi ,\tau ,z \right)}{s_{22}\left( z \right)} \right] =\tilde{C}_ne^{-2i\theta \left( z_{n}^{*} \right)}\mu _{-,1}\left( \chi ,\tau ,z_{n}^{*} \right),\,\, n=1,...,2N,
	\end{align}
\end{subequations}
where $ C_n=\frac{b_n}{s_{11}^{\prime}\left( z_n \right)},\tilde{C}_n=\frac{\tilde{b}_n}{s_{22}^{\prime}\left( z_{n}^{*} \right)}$. 
Therefore, 
\begin{subequations}
	\begin{align}
		&\underset{z=\zeta _n}{\text{Res}}P^-=\left( C_n\left( \zeta _n \right) e^{2i\theta \left( \zeta _n \right)}\mu _{-,2}\left( \chi ,\tau ,\zeta _n \right) ,0 \right),\,\, n=1,...,2N,\\
		&\underset{z=\zeta _{n}^{*}}{\text{Res}}P^+=\left( 0,\tilde{C}_n\left( \zeta _{n}^{*} \right) e^{-2i\theta \left( \zeta _{n}^{*} \right)}\mu _{-,1}\left( \chi ,\tau ,\zeta _{n}^{*} \right) \right),\,\, n=1,...,2N.
	\end{align}
\end{subequations}
When $z \to \infty$, by applying Taylor expansion to Eq.~(\ref{s71}), we get
\begin{align}\label{s35}
	P=I-\frac{1}{z}\left( i\sigma _3\tilde{X}_--\sum_{n=1}^{2N}{\underset{z=\zeta _{n}^{*}}{\text{Res}}P^+}-\sum_{n=1}^{2N}{\underset{z=\zeta _n}{\text{Res}}P^-}-\frac{1}{2\pi i}\int_{\varSigma}{P^+J}d\zeta \right) +O\left( \frac{1}{z^2} \right).
\end{align}
Therefore, one has
\begin{align}\label{s36}
2q_2\left( \chi ,\tau \right)=2\tilde{q}_{2,-}+i\sum_{n=1}^{2N}{\tilde{C}_n\left( \zeta _{n}^{*} \right) e^{-2i\theta \left( \zeta _{n}^{*} \right)}\mu _{-,1}\left( \chi ,\tau ,\zeta _{n}^{*} \right) +\frac{1}{2\pi}\int_{\varSigma}{\left[ P^+J \right] _{12}d\zeta}}.
\end{align}

Next, we present the trace formulae and $\theta$ condition. Recall that discrete spectrum, the functions
\begin{align}\label{s37}
\beta ^-=s_{11}\left( z \right) \prod_{j=1}^{2N}{\frac{z-\zeta _{j}^{*}}{z-\zeta _j}},\,\, \beta ^+=s_{22}\left( z \right) \prod_{j=1}^{2N}{\frac{z-\zeta _j}{z-\zeta _{j}^{*}}}
\end{align}
are analytic in $D^{\mp}$, respectively. Applying Plemelj formulae, one can obtain
$$
\log \beta ^{\pm}=\pm \frac{1}{2\pi i}\int_{\varSigma}{\frac{\log \left[ 1-\rho \left( z \right) \rho ^*\left( z^* \right) \right]}{s-z}ds}, \,\, z\in D^{\pm}.$$
In the special case that $\rho \left( z \right) =\tilde{\rho}\left( z \right) =0$, the scattering coefficients meet
\begin{align}\label{s38}
s_{11}\left( z \right) =\prod_{j=1}^{2N}{\frac{z-\zeta _j}{z-\zeta _{j}^{*}}}, \,\, z\in D^-, \, \, s_{22}\left( z \right) =\prod_{j=1}^{2N}{\frac{z-\zeta _{j}^{*}}{z-\zeta _j}}, \,\, z\in D^+.
\end{align}
And $\theta$ condition can be calculated rapidly:
$$
\text{arg}\left( \frac{\tilde{q}_{2,-}}{\tilde{q}_{2,+}} \right) =4\sum_{j=1}^N{\text{arg}\left( \zeta _j \right)}.
$$

In the special case of reflectionless potential, the reconstruction potential functions can be given by
	\begin{align}\label{s396}
		&q_1\left( \chi ,\tau \right)=\left( \frac{i}{2}\sum_{n=1}^{2N}{\tilde{C}_n}e^{-2i\theta \left( \zeta _{n}^{*} \right)}\mu _{-,11}\left( \chi ,\tau ,\zeta _{n}^{*} \right) \right) _{\tau}^{\frac{1}{2}},\\ \label{s398}
		&q_2\left( \chi ,\tau \right)=\frac{i}{2}\mu _{\pm ,12}^{\left( 1 \right)}=\tilde{q}_{2,-}+\frac{i}{2}\sum_{n=1}^{2N}{\tilde{C}_n}e^{-2i\theta \left( \zeta _{n}^{*} \right)}\mu _{-,11}\left( \chi ,\tau ,\zeta _{n}^{*} \right),
	\end{align}
where
\begin{align}\label{s395}
	\mu _{-,11}\left( \chi ,\tau ,\zeta _{n}^{*} \right) =1-\sum_{j=1}^{2N}{\frac{2i}{\zeta _j}c_j\left( \zeta _{n}^{*} \right) \tilde{q}_{2,-}}+\sum_{j=1}^{2N}{\sum_{k=1}^{2N}{c_j\left( \zeta _{n}^{*} \right) c_{k}^{*}\left( \zeta _{j}^{*} \right) \mu _{-,11}\left( \zeta _{k}^{*} \right)}},
\end{align}
and $c_j(z)=\frac{C_j e^{2i\theta \left( \zeta _j \right)}}{z-\zeta _j}, j=1,...,2N.$
Eq.~(\ref{s395}) can be derived from
\begin{subequations}
	\begin{align} \label{s401}
		&\mu _{-,11}\left( \chi ,\tau ,z \right) =1+\sum_{j=1}^{2N}{\frac{C_je^{2i\theta \left( \zeta _j \right)}\mu _{-,12}\left( \chi ,\tau ,\zeta _j \right)}{z-\zeta _j}},
		\\ \label{s402}
		&\mu _{-,12}\left( \chi ,\tau ,z \right) =-\frac{2i}{z}\tilde{q}_{2,-}+\sum_{k=1}^{2N}{\frac{\tilde{C}_ke^{-2i\theta \left( \zeta _{k}^{*} \right)}\mu _{-,11}\left( \chi ,\tau ,\zeta _{k}^{*} \right)}{z-\zeta _{k}^{*}}.}
	\end{align}
\end{subequations}

\subsection{Riemann-Hilbert problem with double poles}
\hspace{0.7cm} Differing from the above, the discete spectrum is $Z=\left\{ \zeta _n,\hat{\zeta}_n,n=1,...,2N \right\}$, where $\zeta _n=z_n, \, \zeta _{n+N}=\frac{4q_{20}^{2}}{z_{n}^{*}}, \, \hat{\zeta}_n=\frac{4q_{20}^{2}}{z_n}, \, \hat{\zeta}_{n+N}=z_{n}^{*},n=1,...,2N$ are the double zeros of $s_{11}\left( z \right)$ and $s_{22}\left( z \right)$, respectively, meeting $s_{11}\left( \zeta _n \right) =s_{11}^{\prime}\left( \zeta _n \right) =0,s_{22}\left( \hat{\zeta}_n \right) =s_{22}^{\prime}\left( \hat{\zeta}_n \right) =0,n=1,...,2N.$
Similar to the single-pole case, one also has
\begin{align}\label{s41}
	\mu _{+,1}\left( \zeta _n \right) =b_ne^{2i\theta \left( \zeta _n \right)}\mu _{-,2}\left( \zeta _n \right),\,\, \mu _{+,2}\left( \zeta _n \right) =\hat{b}_ne^{-2i\theta \left( \hat{\zeta}_n \right)}\mu _{-,1}\left( \hat{\zeta}_n \right),\,\, n=1,...,2N.
\end{align}
Taking the derivative of Eq.~(\ref{s41}) at points $\zeta_n $ and $\hat{\zeta}_n$, respectively, there are
\begin{subequations}\label{s42}
	\begin{align}
		&\mu _{+,1}^{\prime}\left( \zeta _n \right) =e^{2i\theta \left( \zeta _n \right)}\left[ \left( d_n+2ib_n\theta ^{\prime}\left( \zeta _n \right) \right) \mu _{-,2}+b_n\mu _{-,2}^{\prime}\left( \zeta _n \right) \right],\,\, n=1,...,2N, \\
		&\mu _{+,2}^{\prime}\left( \hat{\zeta}_n \right) =e^{-2i\theta \left( \hat{\zeta}_n \right)}\left[ \left( \hat{d}_n+2i\hat{b}_n\theta ^{\prime}\left( \hat{\zeta}_n \right) \right) \mu _{-,1}+\hat{b}_n\mu _{-,1}^{\prime}\left( \hat{\zeta}_n \right) \right],\,\, n=1,...,2N,
	\end{align}
\end{subequations}
with
$$\hat{b}_n=-\frac{q_{2,-}}{q_{2,+}^{*}}b_n, \, \, \hat{d}_n=d_n\frac{q_{2,-}}{q_{2,+}^{*}}\frac{\zeta _{n}^{2}}{4q_{20}^{2}}.$$
By using Eqs.~(\ref{s41})-(\ref{s42}), one can obtain the following expressions for the coefficients of $\frac{\mu _{+,1}\left( z \right)}{s_{11}\left( z \right)}$ and $ \frac{\mu _{+,2}\left( z \right)}{s_{22}\left( z \right)}$:
\begin{subequations}\label{s431}
	\begin{align}
		&\underset{z=\zeta _n}{P_{-,2}}\left[ \frac{\mu _{+,1}\left( z \right)}{s_{11}\left( z \right)} \right] =\frac{2\mu _{+,1}\left( \zeta _n \right)}{s_{11}^{''}\left( \zeta _n \right)}=A_ne^{2i\theta \left( \zeta _n \right)}\mu _{-,2}\left( \zeta _n \right),
		\\
		&\underset{z=\zeta _n}{\text{Res}}\left[ \frac{\mu _{+,1}\left( z \right)}{s_{11}\left( z \right)} \right] =A_n\left[ e^{2i\theta \left( \zeta _n \right)}\mu _{-,2}^{\prime}\left( \zeta _n \right) +e^{2i\theta \left( \zeta _n \right)}\left( B_n+2i\theta ^{\prime}\left( \zeta _n \right) \right) \mu _{-,2}\left( \zeta _n \right) \right],\\ \label{s432}
		&\underset{z=\hat{\zeta}_n}{P_{-,2}}\left[ \frac{\mu _{+,2}\left( z \right)}{s_{22}\left( z \right)} \right] =\frac{2\mu _{+,2}\left( \hat{\zeta}_n \right)}{s_{22}^{''}\left( \hat{\zeta}_n \right)}=\hat{A}_ne^{-2i\theta \left( \hat{\zeta}_n \right)}\mu _{-,1}\left( \hat{\zeta}_n \right),
		\\
		&\underset{z=\hat{\zeta}_n}{\text{Res}}\left[ \frac{\mu _{+,2}\left( z \right)}{s_{22}\left( z \right)} \right] =\hat{A}_n\left[ e^{-2i\theta \left( \hat{\zeta}_n \right)}\mu _{-,1}^{\prime}\left( \hat{\zeta}_n \right) +e^{-2i\theta \left( \hat{\zeta}_n \right)}\left( \hat{B}_n-2i\theta ^{\prime}\left( \hat{\zeta}_n \right) \right) \mu _{-,1}\left( \hat{\zeta}_n \right) \right],
	\end{align}
\end{subequations}
where 
$$A_n=\frac{2b_n}{s_{11}^{''}\left( \zeta _n \right)},\,\, \hat{A}_n=\frac{2\hat{b}_n}{s_{22}^{''}\left( \hat{\zeta}_n \right)},\,\, B_n=\frac{d_n}{b_n}-\frac{s_{11}^{'''}\left( \zeta _n \right)}{3s_{11}^{''}\left( \zeta _n \right)}, \,\,  \hat{B}_n=\frac{\hat{d}_n}{\hat{b}_n}-\frac{s_{22}^{'''}\left( \hat{\zeta}_n \right)}{3s_{22}^{''}\left( \hat{\zeta}_n \right)}.$$
Construct the sectionally matrix functions $M^\pm$:
\begin{align}
	M^+\left( \chi ,\tau ,z \right) =\left( \mu _{-,1},\frac{\mu _{+,2}}{s_{22}} \right), \, \, M^-\left( \chi ,\tau ,z \right) =\left( \frac{ \mu _{+,1}}{s_{11}},\mu _{-,2}\right),
\end{align}
then the following RH problem can be derived:\\
(1) $M^\pm$ are meromorphic in $\mathbb{C}\setminus \Sigma$, respectively, \\
(2) $M^-\left( \chi ,\tau ,z \right) =M^+\left( \chi ,\tau ,z \right) \left( I-G\left( \chi ,\tau ,z \right) \right), z\in \varSigma $, \\
(3) $M^\pm$ satisfy the residue conditions $\left\{ z:s_{11}\left( z \right) =s_{22}\left( z \right) =s_{11}^{\prime}\left( z \right) =s_{22}^{\prime}\left( z \right) =0 \right\}$ at the zero pionts,\\
(4) $M^{\pm}=I+O\left( \frac{1}{z} \right) $, as $z\rightarrow \infty$, $M^{\pm}=-\frac{i}{z}\sigma _3\tilde{X}_-+O\left( 1 \right)$, as $z\rightarrow 0$,\\
where jump matrix is $$
G\left( \chi ,\tau ,z \right)=\left( \begin{matrix}
	\rho \left( z \right) \tilde{\rho }\left( z \right)&		e^{-2i\theta\left( z \right)}\tilde{\rho} \left( z \right)\\
	-\rho \left( z \right) e^{2i\theta\left( z \right)}&		0\\
\end{matrix} \right) .$$
As in the single-pole case, we solve the RH problem using the Plemelj formula and obtain
\begin{align}\label{s45}
M=I-\frac{i}{z}\sigma _3\tilde{X}_-+\sum_{n=1}^{2N}{\left[ \frac{\underset{z=\hat{\zeta}_n}{\text{Res}}M^+}{z-\hat{\zeta}_n}+\frac{\underset{z=\hat{\zeta}_n}{P_{-,2}}M^+}{\left( z-\hat{\zeta}_n \right) ^2}+\frac{\underset{z=\zeta _n}{\text{Res}}M^-}{z-\zeta _n}+\frac{\underset{z=\zeta _n}{P_{-,2}}M^-}{\left( z-\zeta _n \right) ^2} \right]}+\frac{1}{2\pi i}\int_{\varSigma}{\frac{M^+G}{\zeta -z}}d\zeta. 
\end{align}

Taking the derivative of the second column of Eq.~(\ref{s45}) with respect to $z$ and evaluating the result at point $z=\zeta_{k}$, we obtain
\begin{subequations} \label{s461}
	\begin{align}
		&\mu _{-,2}\left( \zeta _k \right) =\left( \begin{array}{c}
			0\\
			1\\
		\end{array} \right) -\frac{i}{\zeta _k}\left( \begin{array}{c}
			2q_{2,-}\\
			0\\
		\end{array} \right) +\frac{1}{2\pi i}\int_{\varSigma}{\frac{\left( M_+G \right) _2}{s-\zeta _k}ds}+\sum_{n=1}^{2N}{E_n},
		\\ 
		&\mu _{-,2}^{\prime}\left( \zeta _k \right) =\left( \begin{array}{c}
			\frac{2iq_{2,-}}{\zeta _{k}^{2}}\\
			0\\
		\end{array} \right) +\frac{1}{2\pi i}\int_{\varSigma}{\frac{\left( M_+G \right) _2}{\left( s-\zeta _k \right) ^2}ds}-\sum_{n=1}^{2N}{F_n},
	\end{align}
\end{subequations}
where 
$$E_n={\frac{\check{C}_n}{\zeta _k-\hat{\zeta}_n}\mu _{-,1}\left( \hat{\zeta}_n \right) +\check{C}_n\left[ \mu _{-,1}^{\prime}\left( \hat{\zeta}_n \right) +D_n\mu _{-,1}\left( \hat{\zeta}_n \right) \right] },$$ 
$$F_n={ \frac{\check{C}_n}{\zeta _k-\hat{\zeta}_n}\mu _{-,1}^{\prime}\left( \hat{\zeta}_n \right) +\frac{\check{C}_n}{\zeta _k-\hat{\zeta}_n}\left( D_n+\frac{2}{\zeta _k-\hat{\zeta}_n} \right) \mu _{-,1}\left( \hat{\zeta}_n \right) },$$
and
$$\check{C}_n=\frac{\hat{A}_ne^{-2i\theta \left( \hat{\zeta}_n \right)}}{\zeta _k-\hat{\zeta}_n},\,\, D_n=\hat{B}_n-2i\theta ^{\prime}\left( \hat{\zeta}_n \right).$$
Taking the symmetry described by Eq.~(\ref{s271}), one has
\begin{align}\label{s47}
\mu _{-,2}\left( \zeta _k \right) =-\frac{2i}{\zeta _n}q_{2,-}\mu _{-,1}\left( \hat{\zeta}_k \right) , \, \, \mu _{-,2}^{\prime}=\frac{2iq_{2,-}}{\zeta _{k}^{2}}\mu _{-,1}\left( \hat{\zeta}_k \right) +\frac{8iq_{2,-}q_{20}^{2}}{\zeta _{k}^{3}}\mu _{-,1}^{\prime}\left( \hat{\zeta}_k \right).
\end{align}
Substituting the left-hand of Eq.~(\ref{s461}) with Eq.~(\ref{s47}) yields
\begin{subequations}
	\begin{align}
		&\left( \begin{array}{c}
			-\frac{2iq_{2,-}}{\zeta _k}\\
			1\\
		\end{array} \right) +\frac{1}{2\pi i}\int_{\varSigma}{\frac{\left( M_+G \right) _2}{s-\zeta _k}ds}+\sum_{n=1}^{2N}{E_{n}^{\prime}}=0, k=1,...,2N,\\
		&\left( \begin{array}{c}
			\frac{2iq_{2,-}}{\zeta _{k}^{2}}\\
			0\\
		\end{array} \right) +\frac{1}{2\pi i}\int_{\varSigma}{\frac{\left( M_+G \right) _2}{\left( s-\zeta _k \right) ^2}ds}-\sum_{n=1}^{2N}{F_{n}^{\prime}}=0, k=1,...,2N,
	\end{align}
\end{subequations}
where
$$E_{n}^{\prime}={ \check{C}_n\mu _{-,1}^{\prime}\left( \hat{\zeta}_n \right) +\left[ \check{C}_n\left( \frac{1}{\zeta _k-\hat{\zeta}_n}+D_n \right) +\frac{2iq_{2,-}\Delta _{n,k}}{\zeta _k} \right] \mu _{-,1}\left( \hat{\zeta}_n \right) },$$
$$F_{n}^{\prime}={ \left( \frac{\check{C}_n}{\zeta _k-\hat{\zeta}_n}+\frac{8iq_{2,-}q_{20}^{2}\Delta _{n,k}}{\zeta _{k}^{3}} \right) \mu _{-,1}^{\prime}\left( \hat{\zeta}_n \right) +\left[ \frac{\check{C}_n}{\zeta _k-\hat{\zeta}_n}\left( D_n+\frac{2}{\zeta _k-\hat{\zeta}_n} \right) +\frac{2iq_{2,-}\Delta _{n,k}}{\zeta _{k}^{2}} \right] \mu _{-,1}\left( \hat{\zeta}_n \right) }.$$
Then the second harmonic solution can be obtained as
\begin{equation} \label{s48}
	\begin{aligned} 
		&q_2\left( \chi ,\tau \right) =\frac{i}{2}\underset{z\rightarrow \infty}{\lim}\left[ z\mu _{-,12}\left( z \right) \right] \\
		&=q_{2,-}-\frac{1}{4\pi}\int_{\varSigma}{\left( M^+G \right) _{12}d\zeta}+\frac{i}{2}\sum_{n=1}^{2N}{\left\{ \hat{A}_ne^{-2i\theta \left( \hat{\zeta}_n \right)}\left[ \mu _{-,11}^{\prime}\left( \hat{\zeta}_n \right) +\left(\hat{B}_n-2i\theta ^{\prime}\left( \hat{\zeta}_n \right) \right) \mu _{-,11}\left( \hat{\zeta}_n \right) \right] \right\}}.
		\\
	\end{aligned}
\end{equation}
Similarly, in the double-pole case, we define
\begin{align}
	\beta ^-=s_{11}\left( z \right) \prod_{n=1}^{2N}{\frac{\left( z-\hat{\zeta}_n \right) ^2}{\left( z-\zeta _n \right) ^2}}, \, \, \beta ^+=s_{22}\left( z \right) \prod_{n=1}^{2N}{\frac{\left( z-\zeta _n \right) ^2}{\left( z-\hat{\zeta}_n \right) ^2}},
\end{align}
there are the trace formulae:
\begin{subequations}
	\begin{align}
		&s_{11}\left( z \right) =\exp \left[ -\frac{1}{2\pi i}\int_{\varSigma}{\frac{\log \left[ 1-\rho \left( z \right) \rho ^*\left( z^* \right) \right]}{s-z}ds} \right] \prod_{n=1}^{2N}{\frac{\left( z-\zeta _n \right) ^2}{\left( z-\hat{\zeta}_n \right) ^2}},
		\\
		&s_{22}\left( z \right) =\exp \left[ -\frac{1}{2\pi i}\int_{\varSigma}{\frac{\log \left[ 1-\rho \left( z \right) \rho ^*\left( z^* \right) \right]}{s-z}ds} \right] \prod_{n=1}^{2N}{\frac{\left( z-\hat{\zeta}_n \right) ^2}{\left( z-\zeta _n \right) ^2}},
	\end{align}
\end{subequations}
and $\theta$ condition:
$$\text{arg}\left( \frac{q_{2,-}}{q_{2,+}} \right) =8\sum_{j=1}^N{\text{arg}\left( \zeta _j \right)}.$$

	\section{Soliton solutions}
	\subsection{Single-pole solutions}
\hspace{0.7cm}Equivalently, we represent Eq.~(\ref{s395}) in matrix form as follows:
	$$\check{M}Y=\check{B},\,\, \check{M}=I+\check{A},$$ 
where
	$$
	\check{A}=\left( \check{A}_{n,k} \right),\,\, \check{B}=\left( \check{B}_1,......,\check{B}_{2N} \right) ^{\top},\,\, Y=\left( Y_1,......,Y_{2N} \right) ^{\top}
	$$ 
with $n,k = 1,...,2N,$ 
	$$
	\check{A}_{n,k}=\sum_{j=1}^{2N}{c_j\left( \zeta _{n}^{*} \right) c_{k}^{*}\left( \zeta _{j}^{*} \right)},\,\, \check{B}_n=1-2i\tilde{q}_{2,-}\sum_{j=1}^{2N}{\frac{c_j\left( \zeta _{n}^{*} \right)}{\zeta _j}},\,\, Y_n=\mu _{-,11}\left( \zeta _{n}^{*} \right),
	$$ 
then one can obtain
	$$
	Y_n=\frac{\det M_{n}^{rep}}{\det \check{M}},\,\, M_{n}^{rep}=\left( \check{M}_1,...,\check{M}_{n-1},\check{B},\check{M}_{n+1}...,\check{M}_{2N} \right),$$
thus we rewrite the potiential function as
	\begin{align}\label{s51}
		&q_1\left( \chi ,\tau \right)=\left( \frac{i}{2}\frac{\det M^{aug}}{\det \check{M}} \right) _{\tau}^{\frac{1}{2}},\\ \label{s52}
		&q_2\left( \chi ,\tau \right)=\tilde{q}_{2,-}+\frac{i}{2}\sum_{n=1}^{2N}{Z_n\frac{\det M_{n}^{rep}}{\det \check{M}}}=\tilde{q}_{2,-}+\frac{i}{2}\frac{\det M^{aug}}{\det \check{M}},
	\end{align}
with $$Z_n=\tilde{C}_ne^{-2i\theta \left( \zeta _{n}^{*} \right)},\,\, M^{aug}=\left( \begin{matrix}
	0&		Z^{\top}\\
	B&		\check{M}\\
\end{matrix} \right).$$

\begin{figure}[H]
		\centering
		\begin{subfigure}[b]{0.23\textwidth}
			\centering
			\includegraphics[width=\textwidth]{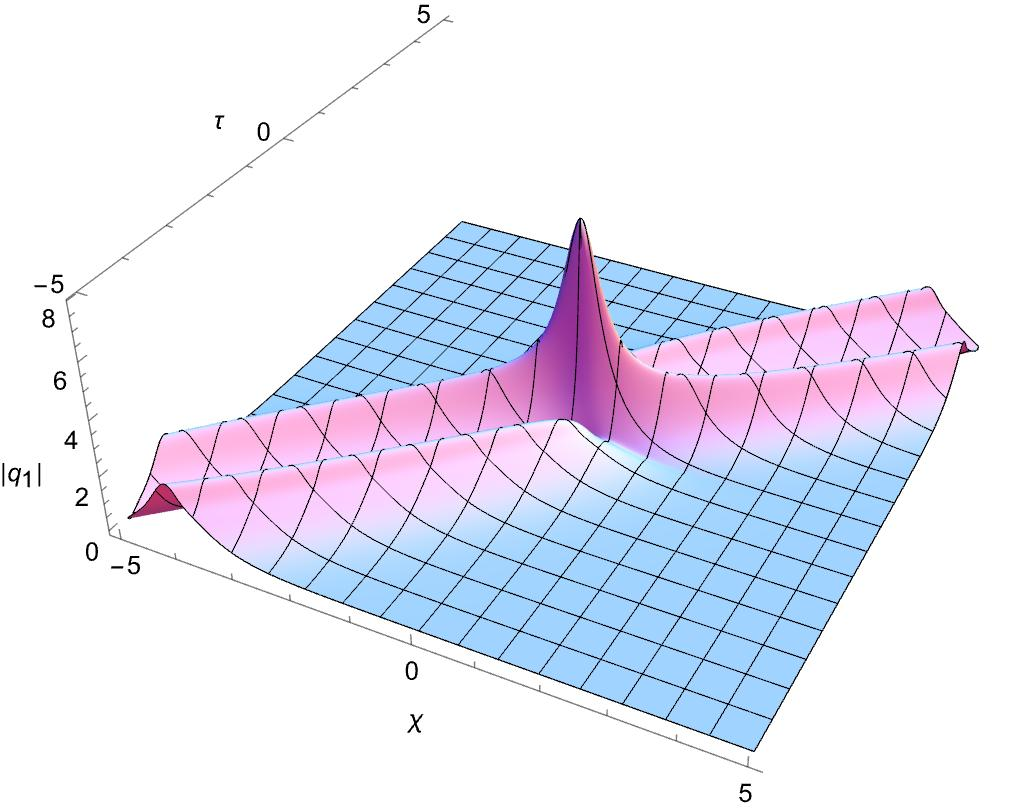}
			\caption{}
		\end{subfigure}
		\hfill
		\begin{subfigure}[b]{0.23\textwidth}
			\centering
			\includegraphics[width=\textwidth]{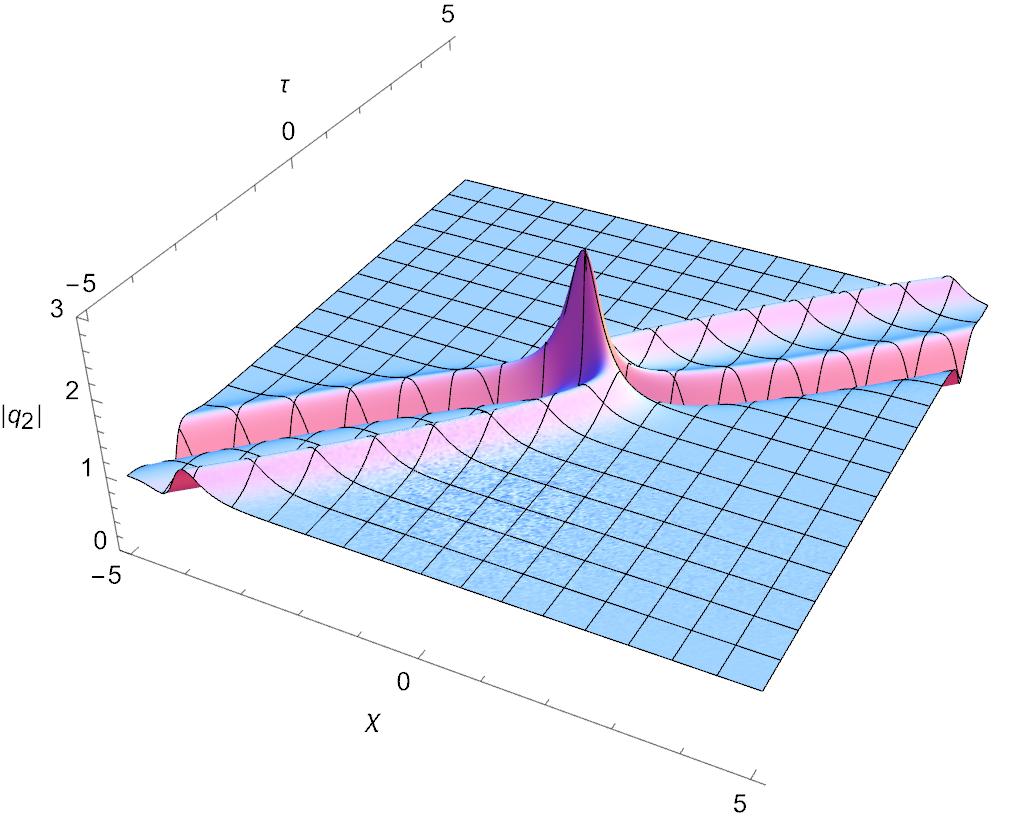}
		    \caption{}
		\end{subfigure}
		\hfill
		\begin{subfigure}[b]{0.23\textwidth}
			\centering
			\includegraphics[width=\textwidth]{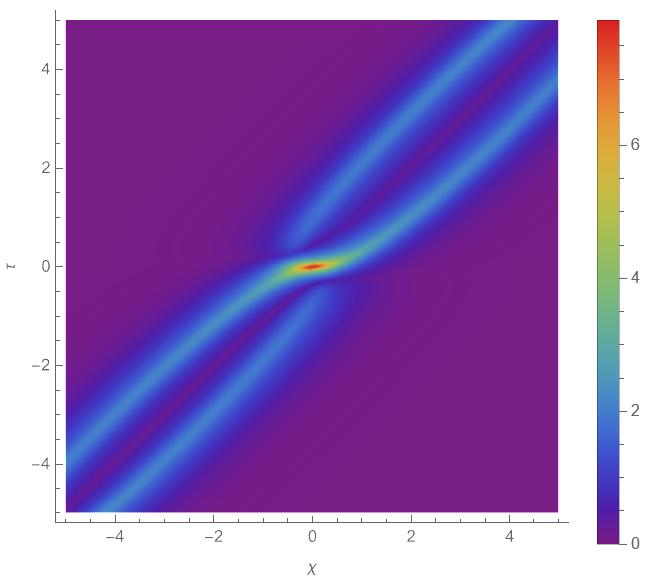}
			\caption{}
		\end{subfigure}
		\hfill
		\begin{subfigure}[b]{0.23\textwidth}
			\centering
			\includegraphics[width=\textwidth]{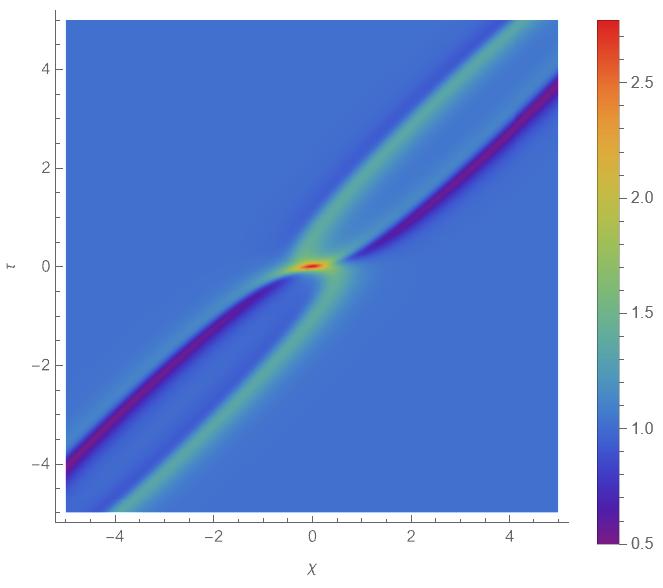}
			\caption{}
		\end{subfigure}
		\caption{Two-soliton solutions with $\tilde{q}_{2,-}=1,\ \zeta_1=2i+\frac{1}{2}$. (a) The three-dimension plot of $q_1$, (b) the three-dimension plot of $q_2$, (c) the density map of $q_1$, (d) the density map of $q_2$.}
		\label{f1}
	\end{figure}

\begin{figure}[H]
		\centering
		\begin{subfigure}[b]{0.23\textwidth}
			\centering
			\includegraphics[width=\textwidth]{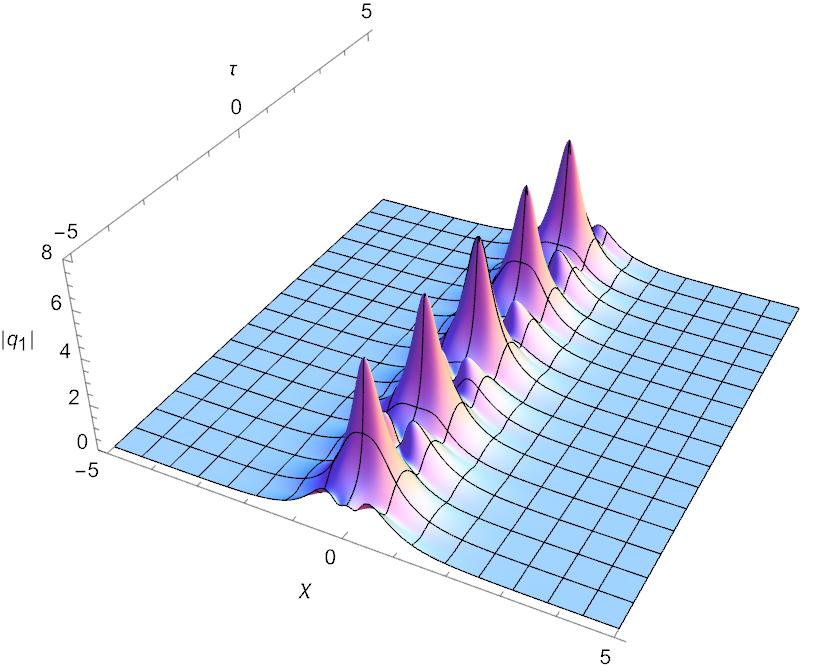}
			\caption{}
		\end{subfigure}
		\hfill
		\begin{subfigure}[b]{0.23\textwidth}
			\centering
			\includegraphics[width=\textwidth]{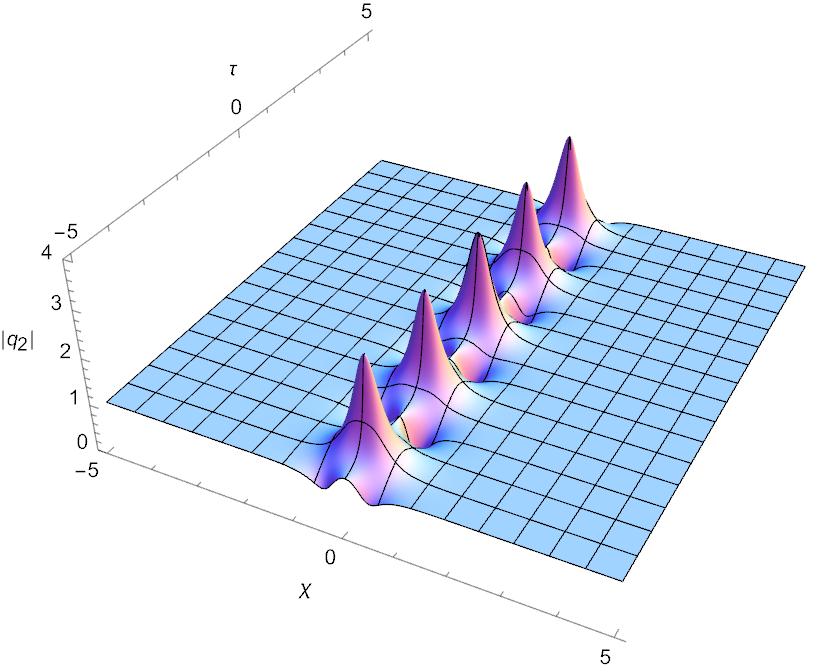}
			\caption{}
		\end{subfigure}
		\hfill
		\begin{subfigure}[b]{0.23\textwidth}
			\centering
			\includegraphics[width=\textwidth]{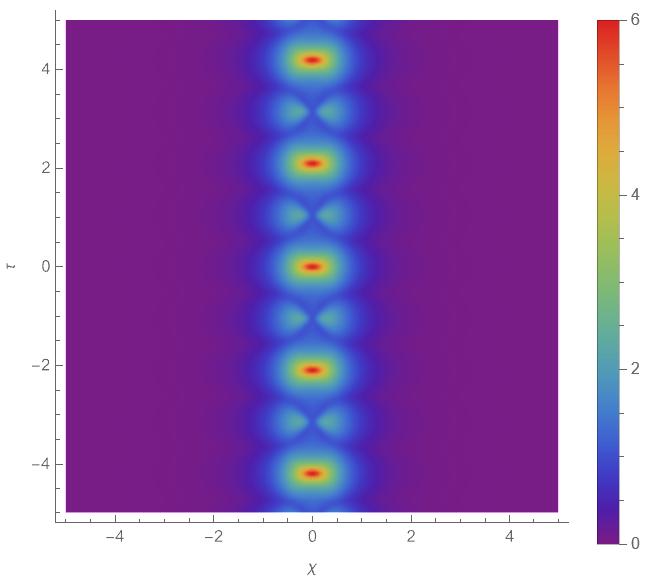}
			\caption{}
		\end{subfigure}
		\hfill
		\begin{subfigure}[b]{0.23\textwidth}
			\centering
			\includegraphics[width=\textwidth]{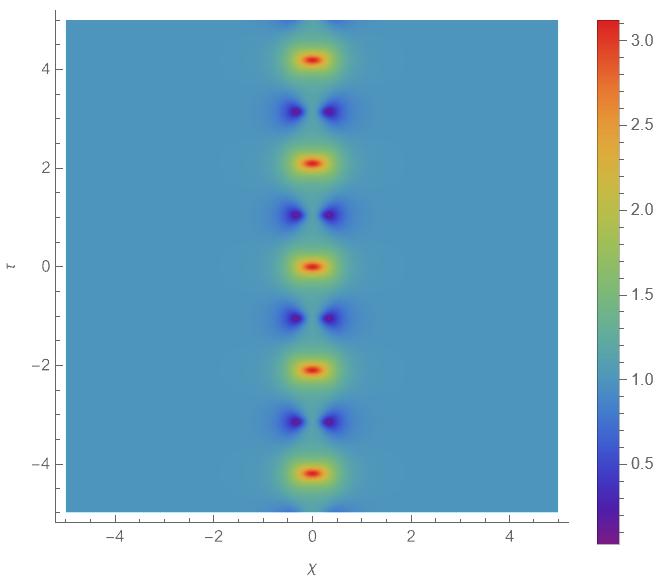}
			\caption{}
		\end{subfigure}
		\caption{Breather solutions with $\tilde{q}_{2,-}=1,\ \zeta_{1}=2\sqrt{2}i$. (a) The three-dimension plot of $q_1$, (b) the three-dimension plot of $q_2$, (c) the density map of $q_1$, (d) the density map of $q_2$.}
		\label{f2}
	\end{figure}

Based on Eqs.~(\ref{s51})-(\ref{s52}), we discuss the influence of different parameter assignments on the dynamic behavior of the virtual solutions. Without loss of generality, we set $\tilde{q}_{2,-}=1$. For $N=1$, one can assume that $\zeta_1=2i+\frac{1}{2}$, there are two solitons collide with each other shown in Fig.~\ref{f1}. However, when the values $\zeta_{1}=2\sqrt{2}i$ are taken, the breather solutions are obtained and shown in Fig.~\ref{f2}. We can find that the discrete spectral points located on the imaginary axis may correspond to the breather solutions.

\begin{figure}[H]
	\centering
	\begin{subfigure}[b]{0.23\textwidth}
		\centering
		\includegraphics[width=\textwidth]{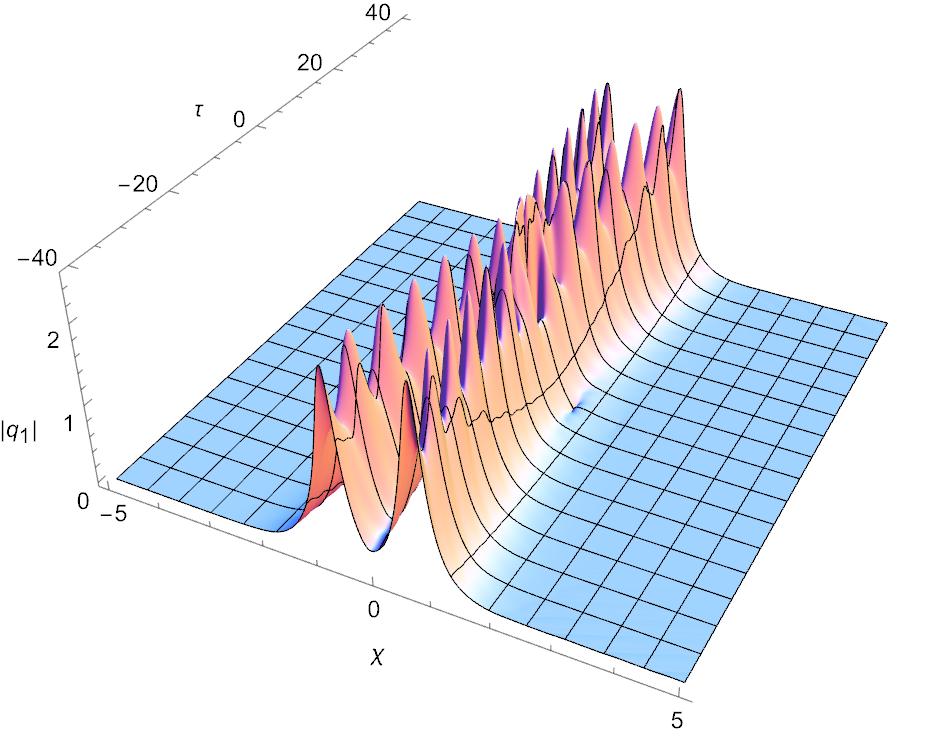}
		\caption{}
	\end{subfigure}
	\hfill
	\begin{subfigure}[b]{0.23\textwidth}
		\centering
		\includegraphics[width=\textwidth]{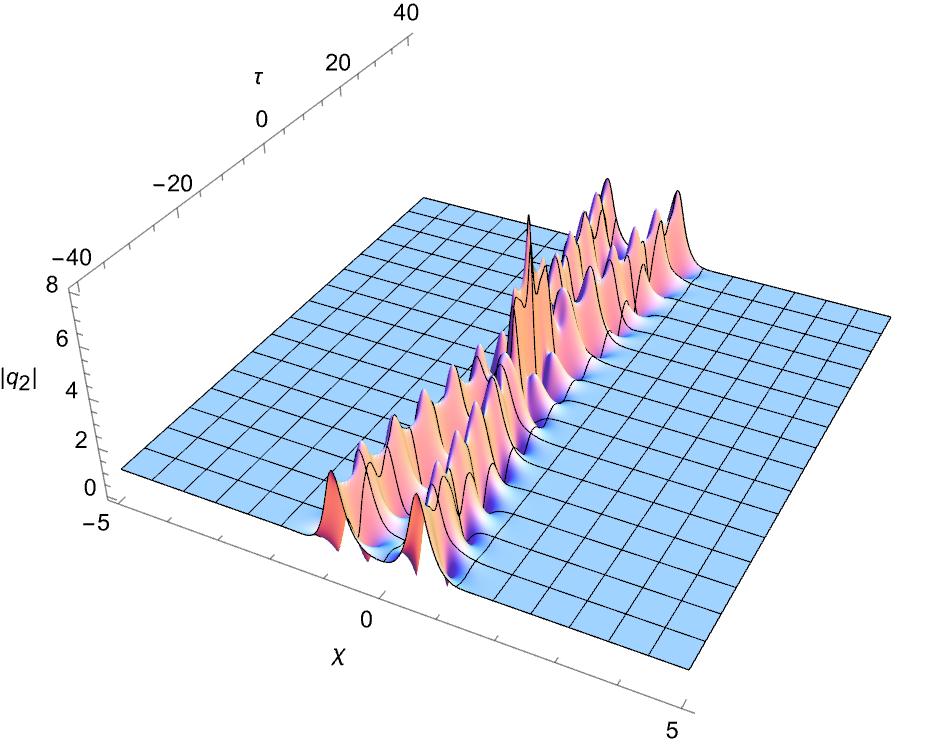}
		\caption{}
	\end{subfigure}
	\hfill
	\begin{subfigure}[b]{0.23\textwidth}
		\centering
		\includegraphics[width=\textwidth]{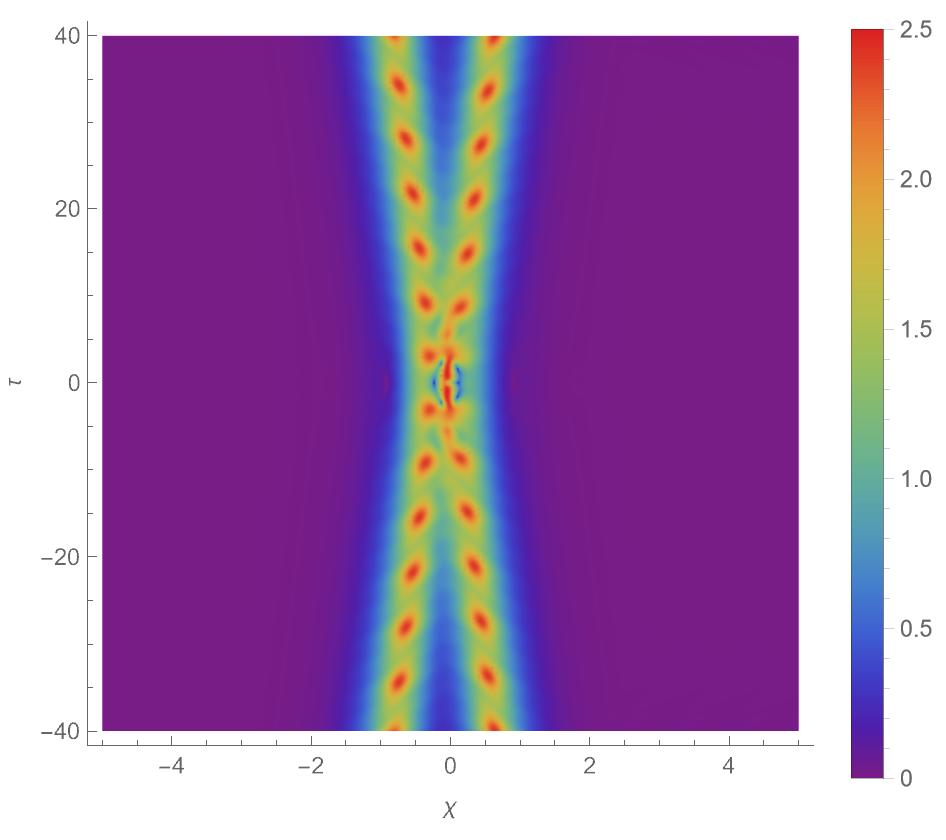}
		\caption{}
	\end{subfigure}
	\hfill
	\begin{subfigure}[b]{0.23\textwidth}
		\centering
		\includegraphics[width=\textwidth]{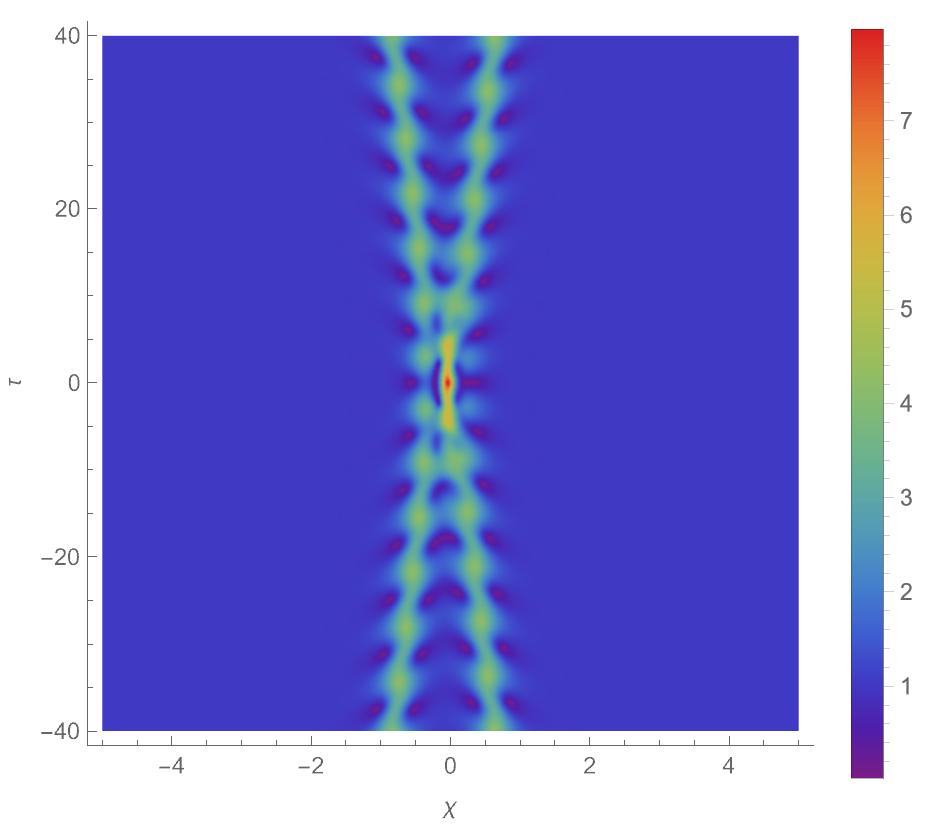}
		\caption{}
	\end{subfigure}
	\caption{Two-breather solutions with $\tilde{q}_{2,-}=1,\ \zeta_1=\frac{1}{4}-\frac{\sqrt{3}}{4}i,\zeta_2=-\frac{1}{4}+\frac{\sqrt{3}}{4}i$. (a) The three-dimension plot of $q_1$, (b) the three-dimension plot of $q_2$, (c) the density map of $q_1$, (d) the density map of $q_2$.}
	\label{f3}
\end{figure}

\begin{figure}[H]
	\centering
	\begin{subfigure}[b]{0.23\textwidth}
		\centering
		\includegraphics[width=\textwidth]{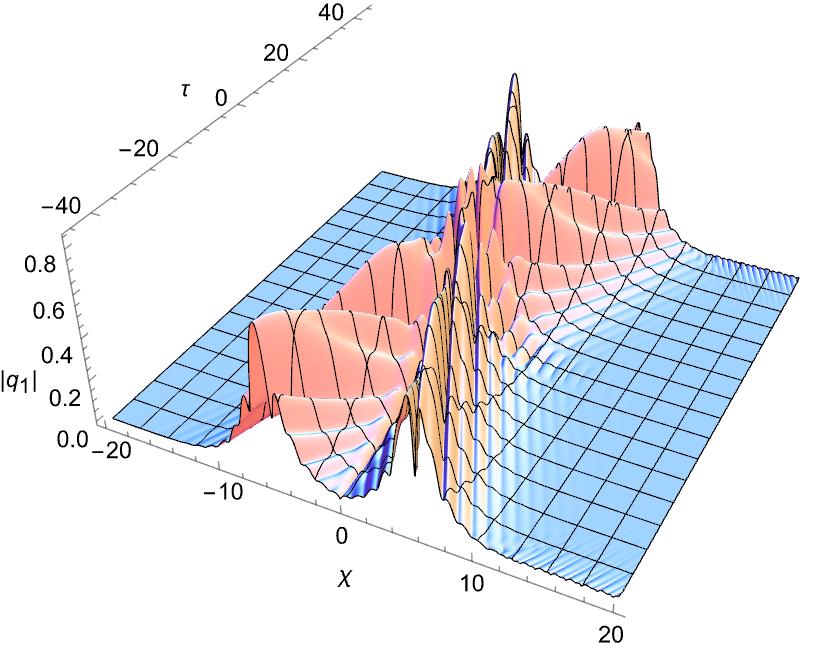}
		\caption{}
	\end{subfigure}
	\hfill
	\begin{subfigure}[b]{0.23\textwidth}
		\centering
		\includegraphics[width=\textwidth]{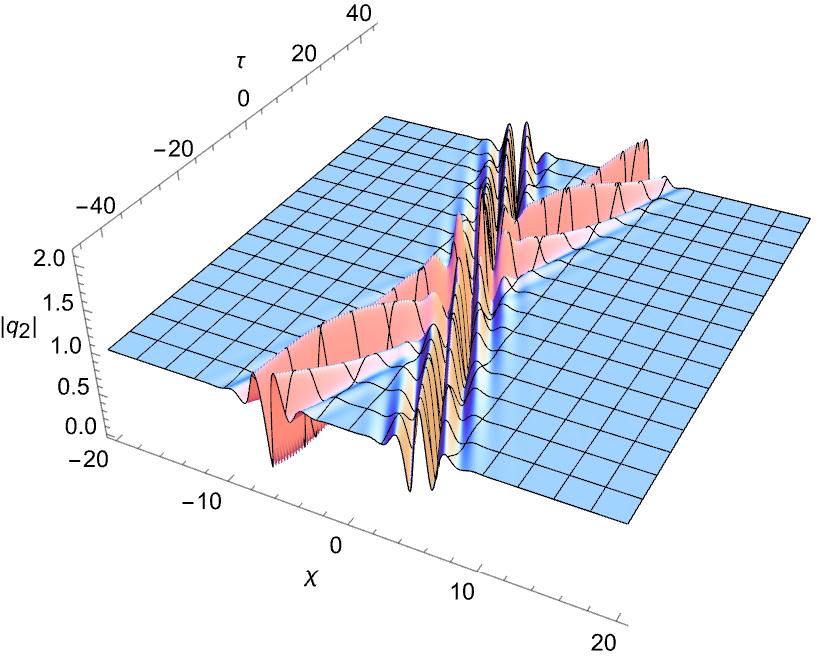}
		\caption{}
	\end{subfigure}
	\hfill
	\begin{subfigure}[b]{0.23\textwidth}
		\centering
		\includegraphics[width=\textwidth]{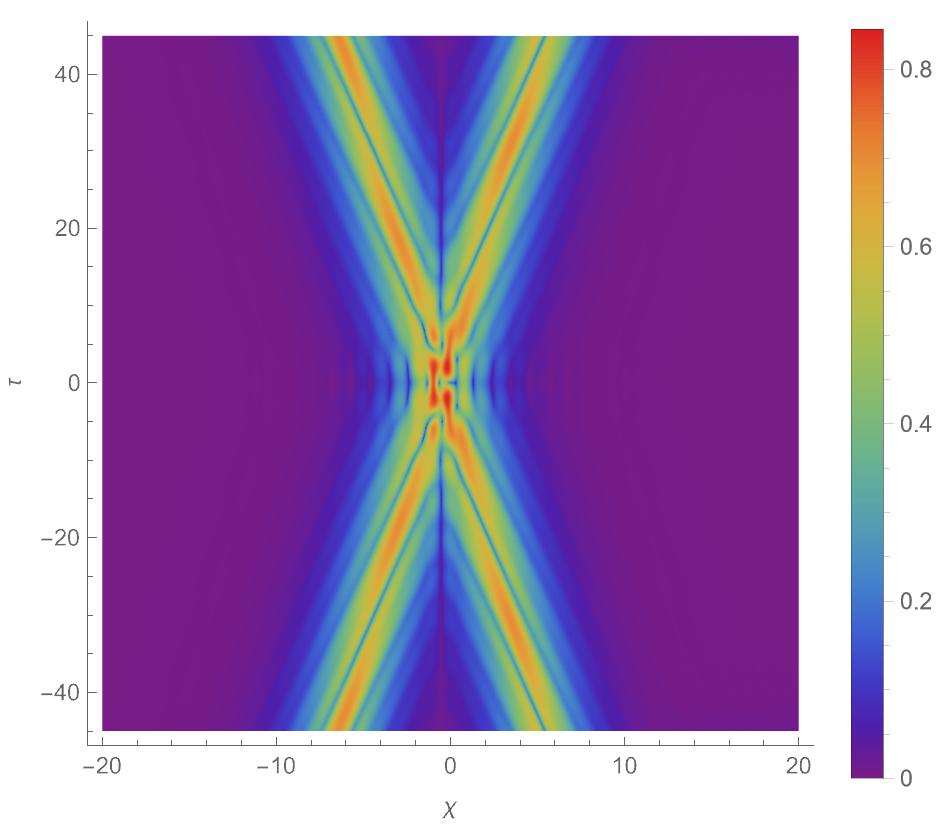}
		\caption{}
	\end{subfigure}
	\hfill
	\begin{subfigure}[b]{0.23\textwidth}
		\centering
		\includegraphics[width=\textwidth]{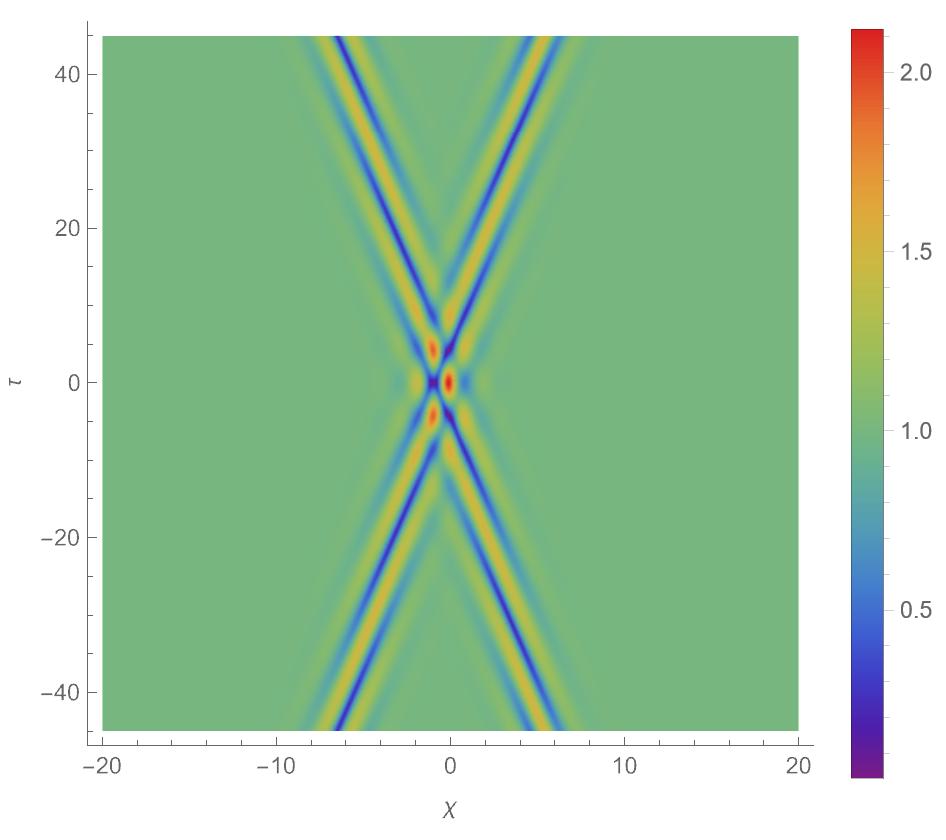}
		\caption{}
	\end{subfigure}
	\caption{Multi-soliton solutions with $\tilde{q}_{2,-}=1,\ \zeta _1=e^{-\frac{\pi}{12}i},\zeta _2=e^{\frac{13\pi}{12}i}$. (a) The three-dimension plot of $q_1$, (b) the three-dimension plot of $q_2$, (c) the density map of $q_1$, (d) the density map of $q_2$.}
	\label{f4}
\end{figure}

Similarly, for $N=2$, in the case of $\zeta_1=\frac{1}{4}-\frac{\sqrt{3}}{4}i$ and $\zeta_2=-\frac{1}{4}+\frac{\sqrt{3}}{4}i$, there are two breathers colliding with each other in Fig.~\ref{f3}. When $\zeta_1=e^{-\frac{\pi}{12}i}$ and $\zeta_2=e^{\frac{13\pi}{12}i}$ are taken, multi-soliton solutions are presented in Fig.~\ref{f4}.

\subsection{Double-pole solutions}
	\hspace{0.7cm} In the reflectionless case, according to Eq.~(\ref{s48}), one can solve with matrices and obtain
	\begin{align}
		UX=\check{V},
	\end{align}
where
    $$
	X_n=\mu _{-,11}\left( \hat{\zeta}_n \right),\,\, X_{2N+n}=\mu _{-,11}^{\prime}\left( \hat{\zeta}_n \right),\,\, \check{V}_k=\frac{2iq_{2,-}}{\zeta _k},\,\, \check{V}_{2N+k}=\frac{2iq_{2,-}}{\zeta _{k}^{2}},
	$$ $$
	U_{k,n}=C_n\left( \zeta _k \right) \left( \frac{1}{\zeta _k-\hat{\zeta}_n}+D_n \right) +\frac{2iq_{2,-}\Delta _{kn}}{\zeta _k},\,\, U_{k,2N+n}=C_n\left( \zeta _k \right),
	$$ $$
U_{2N+k,n}=\frac{C_n\left( \zeta _k \right)}{\zeta _k-\hat{\zeta}_n}\left( \frac{2}{\zeta _k-\hat{\zeta}_n}+D_n \right) +\frac{2iq_{2,-}\Delta _{kn}}{\zeta _{k}^{2}},\,\, U_{2N+k,2N+n}=\frac{C_n\left( \zeta _k \right)}{\zeta _k-\hat{\zeta}_n}+\frac{8iq_{2,-}q_{20}^{2}\Delta _{kn}}{\zeta _{k}^{3}},
	$$ 
with $k,n = 1,...,2N$. Then we can get
	\begin{align} \label{s53}
&q_1\left( \chi ,\tau \right) =\left( \frac{i}{2}\frac{\det U^{aug}}{\det U} \right) _{\tau}^{\frac{1}{2}},\\ \label{s54}
&q_2\left( \chi ,\tau \right) =q_{2,-}+\frac{i}{2}\frac{\det U^{aug}}{\det U},
	\end{align}
$$
U^{aug}=\left( \begin{matrix}
	0&		\hat{A}_{n}^{\prime}\\
	V&		U\\
\end{matrix} \right),\,\, \hat{A}_{n}^{\prime}=\left( \hat{B}_1-2i\theta ^{\prime}\left( \hat{\zeta}_1 \right) ,...,\hat{B}_{2N}-2i\theta ^{\prime}\left( \hat{\zeta}_{2N} \right) ,\hat{A}_1e^{-2i\theta \left( \hat{\zeta}_1 \right)},...,\hat{A}_{2N}e^{-2i\theta \left( \hat{\zeta}_{2N} \right)} \right).
$$

\begin{figure}[H]
		\centering
		\begin{subfigure}[b]{0.23\textwidth}
			\centering
			\includegraphics[width=\textwidth]{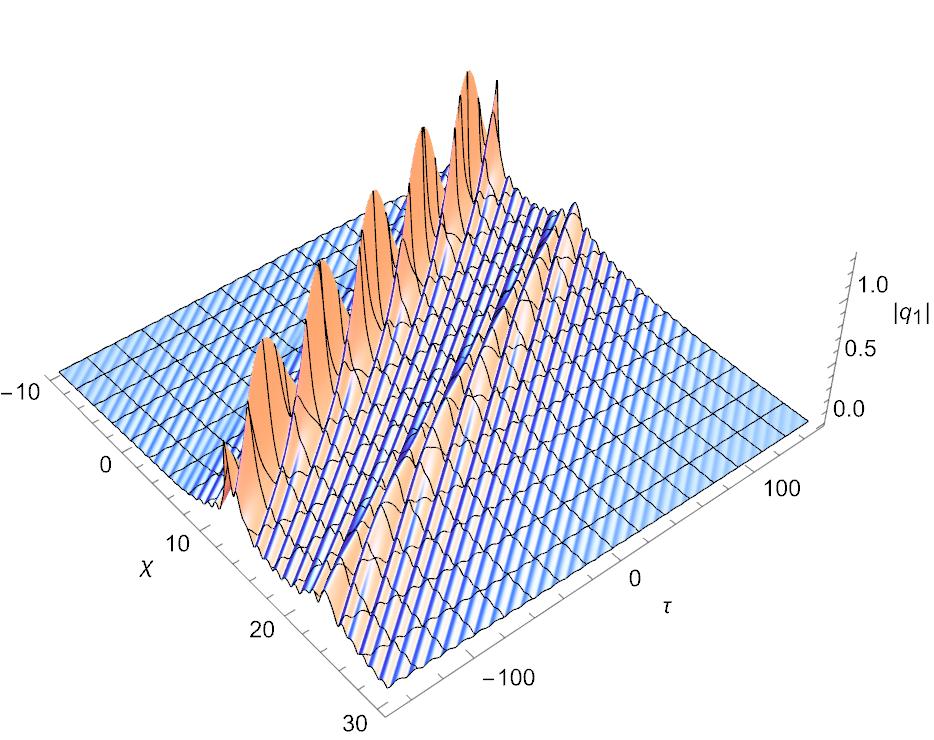}
			\caption{}
		\end{subfigure}
		\hfill
		\begin{subfigure}[b]{0.23\textwidth}
			\centering
			\includegraphics[width=\textwidth]{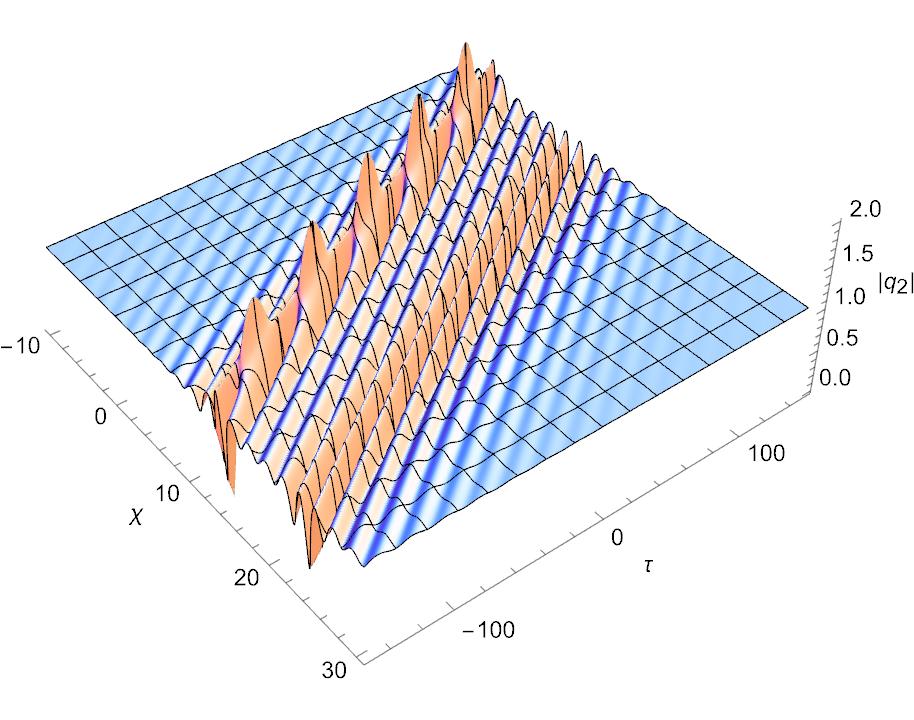}
			\caption{}
		\end{subfigure}
		\hfill
		\begin{subfigure}[b]{0.23\textwidth}
			\centering
			\includegraphics[width=\textwidth]{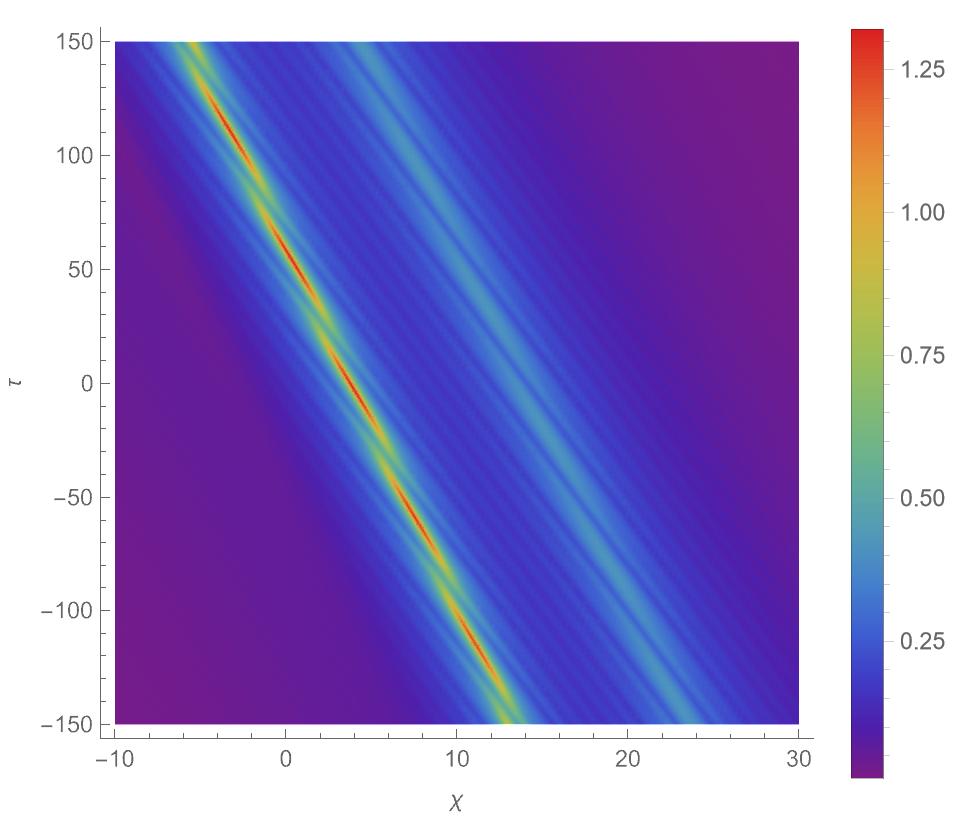}
			\caption{}
		\end{subfigure}
		\hfill
		\begin{subfigure}[b]{0.23\textwidth}
			\centering
			\includegraphics[width=\textwidth]{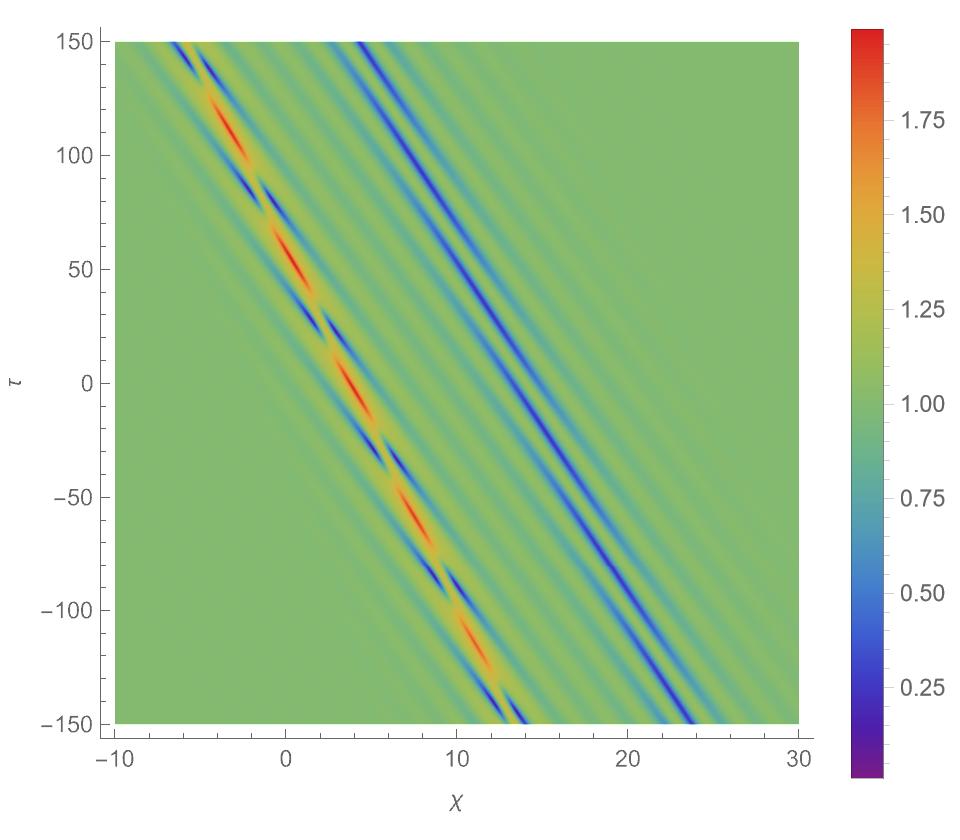}
			\caption{}
		\end{subfigure}
		\caption{Two-soliton and breather-soliton solution with $q_{2,-}=1,\ \alpha = 0.5,\ \zeta_1=e^{-\frac{1}{48}\pi i}$. (a) The three-dimension plot of $q_1$, (b) the three-dimension plot of $q_2$, (c) the density map of $q_1$, (d) the density map of $q_2$.}
		\label{f5}
	\end{figure}

\begin{figure}[H]
		\centering
		\begin{subfigure}[b]{0.23\textwidth}
			\centering
			\includegraphics[width=\textwidth]{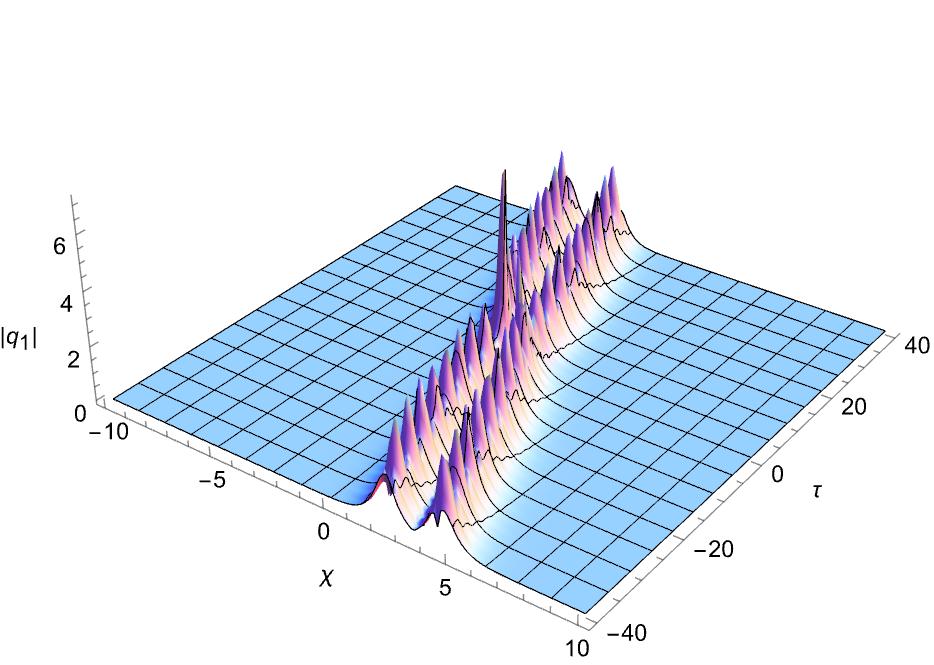}
			\caption{}
		\end{subfigure}
		\hfill
		\begin{subfigure}[b]{0.23\textwidth}
			\centering
			\includegraphics[width=\textwidth]{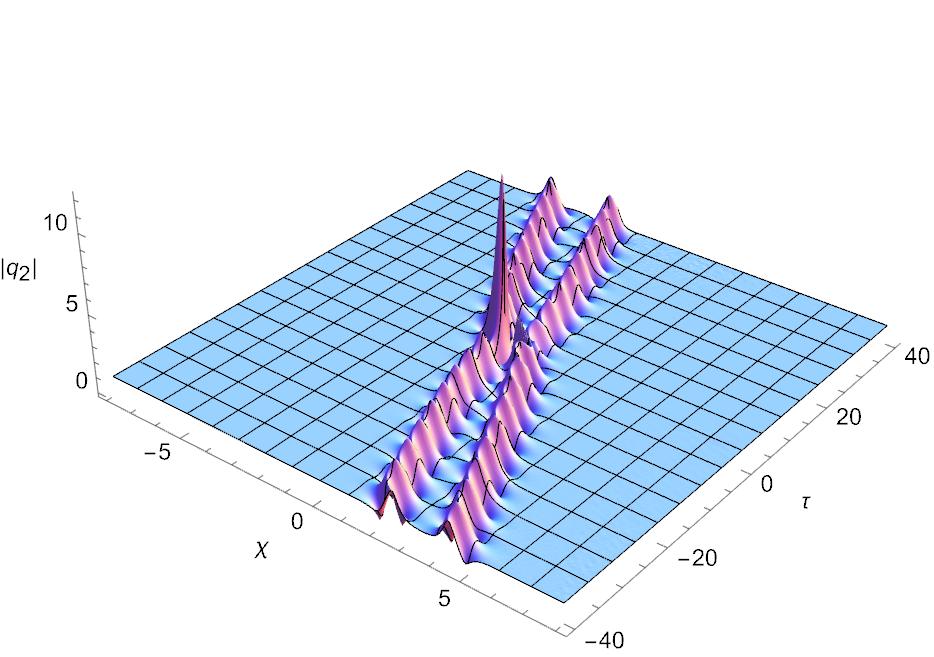}
			\caption{}
		\end{subfigure}
		\hfill
		\begin{subfigure}[b]{0.23\textwidth}
			\centering
			\includegraphics[width=\textwidth]{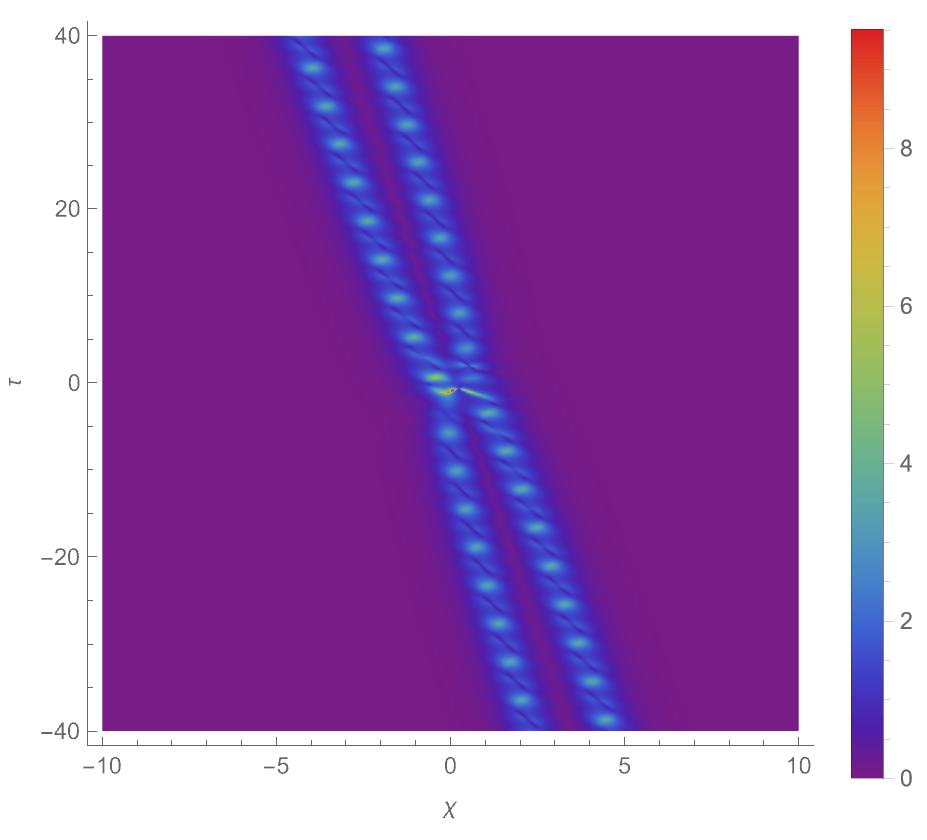}
			\caption{}
		\end{subfigure}
		\hfill
		\begin{subfigure}[b]{0.23\textwidth}
			\centering
			\includegraphics[width=\textwidth]{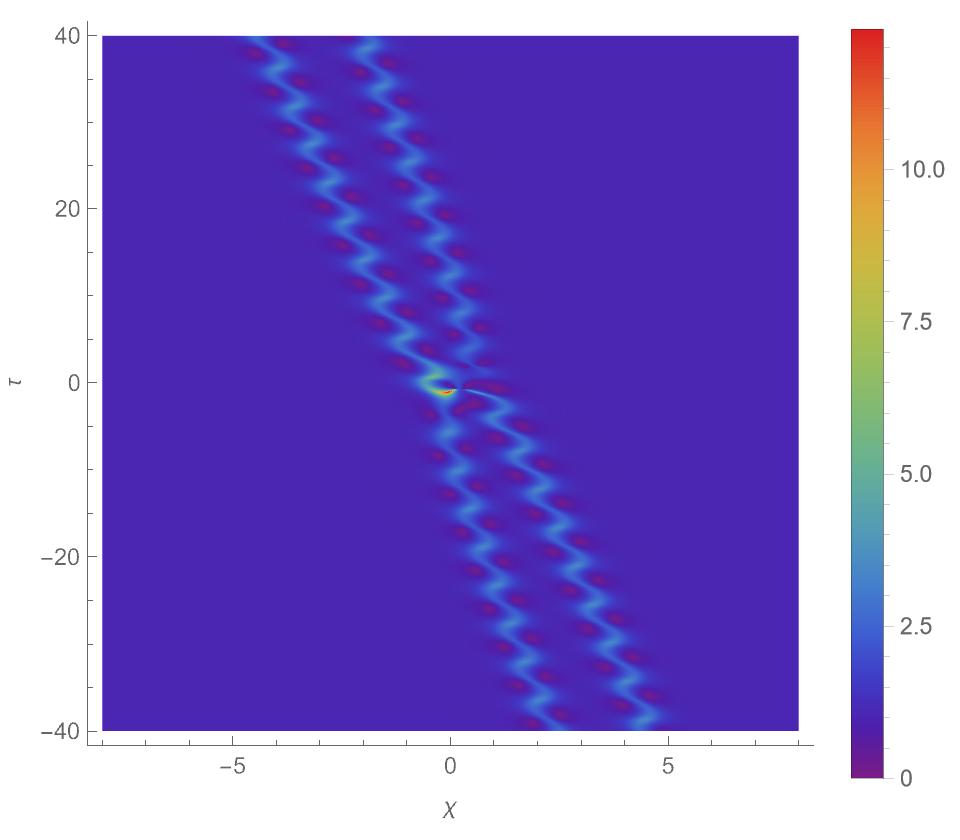}
			\caption{}
		\end{subfigure}
		\caption{Two-breather solution with $q_{2,-}=1,\ \alpha=1,\ \zeta_1=e^{-\frac{5}{12}\pi i}$. (a) The three-dimension plot of $q_1$, (b) the three-dimension plot of $q_2$, (c) the density map of $q_1$, (d) the density map of $q_2$.}
		\label{f6}
	\end{figure}

	According to Eqs.~(\ref{s53})-(\ref{s54}), we just discuss the dynamic behavior of solutions in the case of $N=1$. When $q_{2,-}=1$, $\alpha = 0.5$, and $\zeta_1=e^{-\frac{1}{48}\pi i}$ are taken as shown in Fig.~\ref{f5}, there are two solitons that propagate in the same direction for $q_1$ $\left(\text{Fig.~5a}\right)$, and two breathers propagate in the same direction for $q_2$ $\left(\text{Fig.~5b}\right)$. In Fig.~\ref{f6}, when we take $q_{2,-}=1$, $\alpha=1$, and $\zeta_1=e^{-\frac{5}{12}\pi i}$, the images of $q_1$ and $q_2$ both show that the two breathers move towards each other and collide.
	
    \section{Breather solution asymptotics analysis}
    \subsection{One-breather asymptotics}
    \hspace{0.7cm} Define $\zeta_{n}=iz_ne^{i\alpha_n},$  which satisfies symmetries $\zeta_{n+N}=\frac{1}{\zeta _{n}^{*}}$. After a simple calculation, Eq.~(\ref{s52}) can be written as
    \begin{align}\label{s61}
    	q_2\left( \chi,\tau \right)=q_{2,-}e^{i\alpha _s}\frac{\det \left( \begin{matrix}
    			\left( I-\textbf{1}_{2N}L^* \right) X^*&		-\Gamma L\\
    			\Gamma ^*L^*&		X^{-1}\\
    		\end{matrix} \right)}{\det \left( \begin{matrix}
    			I&		-\Gamma L\\
    			\Gamma ^*L^*&		I\\
    		\end{matrix} \right)},
    \end{align}
    where $\alpha_{s}=-4\sum_{j=1}^{N}{\alpha_{j}}$, $I$ is the $2N \times 2N$ identity matrix, $\textbf{1}_{2N} = \textbf{1}\otimes \textbf{1}$ with $\textbf{1}= \left( 1,1,\cdots ,1,1 \right) ^{\top}$ and $\otimes$ denotes outer product. Besides,
    $$L\left( \chi ,\tau \right) =\mathrm{diag}\left( l_1\left( \chi ,\tau \right) ,l_2\left( \chi ,\tau \right) ,\cdots ,l_{2N-1}\left( \chi ,\tau \right) ,l_{2N}\left( \chi ,\tau \right) \right) 
    ,$$ 
    $$X =\mathrm{diag}\left( \frac{1}{\zeta_1},\frac{1}{\zeta_2},\cdots ,\frac{1}{\zeta_{2N-1}},\frac{1}{\zeta_{2N}} \right),\,\, \Gamma =\left(\gamma_{ij} \right) _{2N\times 2N},$$
    with $\gamma_{i,j}=\frac{1}{\zeta_{i}^{*}-\zeta_{j}}, l_{j}\left( \chi ,\tau \right) = e^{\tilde{\chi}_j\left( \chi ,\tau \right) +is_{j}\left( \chi ,\tau \right)}$ for $i,j = 1,\dots,2N$, and
    $$\tilde{\chi}_j\left( \chi ,\tau \right) =-c_{j,+}\cos\alpha_{j}\chi+\frac{2\sin 2\alpha_{j}}{d_{j,+}-2\cos 2\alpha_{j}}\alpha\tau+\zeta_{j},
    $$
    $$s_j\left( \chi ,\tau \right) =\,\,-c_{j,-}\sin\alpha_{j}\chi+\frac{d_{j,-}}{d_{j,+}-2\cos 2\alpha_{j}}\alpha\tau+\phi_{j},
    $$
    $$c_{j,\pm}=z_{j}\pm \frac{1}{z_{j}},\,\, d_{j,\pm}=z_{j}^{2}\pm \frac{1}{z_{j}^{2}},$$
    for clarity, the transformed solution will be denoted by $q_{2}^{1}$ instead of $q_{2}$ in the following text. 
    
    For $N=1$, one can obtain breather solutions $q_{2}^{1}$ corresponding to velocity $V_1$ and discrete spectrum $\zeta_1$. When $\tau \to \pm  \infty$ we analyze the asymptotic behavior of the solution along the line $\chi=V\tau +y$. Substituting $\chi=V\tau +y$ into Eq.~(\ref{s61}), it is easy to see that depending on the different values of $V$, $\left| l_1 \right|$ and $\left| l_2 \right|$ have two cases in the limit $\tau \to \pm \infty$: both tend to infinity or both decay to zero. The specific analysis is as follows:
    
    If both $\left| l_1 \right|$ and $\left| l_2 \right|$ tend to zero, $L$ and $L^*$ also tend to the zero matrix, then the solution becomes 
    $$q_{2}^{1}\left( \chi,\tau \right)=q_{2,-}e^{-4i\alpha _1}\frac{\det \left( \begin{matrix}
    		X^*&		0\\
    		0&		X^{-1}\\
    	\end{matrix} \right)}{\det \left( \begin{matrix}
    		I&		0\\
    		0&		I\\
    	\end{matrix} \right)} = q_{2,-} + o\left( 1 \right).$$
    
    If both $\left| l_1 \right|$ and $\left| l_2 \right|$ tend to infinity, we multiply the numerator and denominator of Eq.~(\ref{s61}) by the inverse of $\mathrm{diag}\left(l_{1}^{*},l_{2}^{*},l_1,l_2 \right)$ from the right. Then, in this limit the solution becomes 
    $$q_{2}^{1}\left( \chi,\tau \right)=q_{2,-}\,\,e^{i\alpha _s}\frac{\det \left( \begin{matrix}
    		-1_{2N}X^*&		-\Gamma\\
    		\Gamma ^*&		0\\
    	\end{matrix} \right)}{\det \left( \begin{matrix}
    		0&		-\Gamma\\
    		\Gamma ^*&		0\\
    	\end{matrix} \right)}=q_{2,-}e^{-4i\alpha _1} + o\left( 1 \right).$$
    Introducing $V_1=\frac{2\alpha \sin 2\alpha _1\,\,}{\left( d_{+,1}-2\cos 2\alpha _1 \right) c_{+,1}\cos \alpha _1}$, when $\left| \tau \right| \to \infty$, an example is shown in Fig.~7(a). 
    
    \subsection{Breather-breather interactions}
    \hspace{0.7cm} Now we discuss the breather interactions in the framework of two-breather solutions. Similar to the one-breather case, we will observe the different asymptotic behaviors of the parameters $l_j$. Without loss of generality, the values of discrete eigenvalues are taken such that $V_1 < V_2$. Due to symmetry, introducing the notation $l_{j,j+2}$ to represent $l_j$ and $l_{j+2}$ for $j=1,2$. One can substitute $\chi=V\tau +y$ into the exact solution Eq.~(\ref{s61}). It is easy to see that depending on the different values of the $V$, five cases should be taken into account:
1) $V<V_1$, all $l_{j}$ grow/decay as $\tau \to \pm \infty$, respectively;
2) $V=V_1$, $l_{1,3}$ is always bounded, while $l_{2,4}$ grows/decays as $\tau \to \pm \infty$;
3) $V_1<V<V_2$, $l_{1,3}$ decays/grows as $\tau \to \pm \infty$, $l_{2,4}$ is the opposite;
4) $V=V_2$, $l_{2,4}$ is always bounded, while $l_{1,3}$ decays/grows as $\tau \to \pm \infty$;
5) $V>V_2$, all $l_{j}$ decay/grow as $\tau \to \pm \infty$, respectively.

For simplicity, we only present the discussion of the following five cases here: $\text{a}$. $l_j$ decay; $\text{b}$. $l_j$ grow; $\text{c}$. $l_{1,3}$ decays, $l_{2,4}$ is bounded; $\text{d}$. $l_{1,3}$ grows, $l_{2,4}$ decays; $\text{e}$. $l_{1,3}$ grows and $l_{2,4}$ is bounded. The detailed analysis is as follows:\\
$(\text{a})$ When all $l_j$ decay, $L$ and $L^*$ tend to the zero matrix, hence, the breather sloution will reduce to $q_{2,-}$ in this limit.\\
$(\text{b})$ When all $l_j$ grow, for the convenience of analysing, one can multiply the numerator and denominator of Eq.~(\ref{s61}) by the inverse of $\mathrm{diag}\left( l_{1}^{*},l_{2}^{*},l_{3}^{*},l_{4}^{*},l_1,l_2,l_3,l_4 \right)$ from the right, then one can obtain $$q_2\left( \chi,\tau \right)=q_{2,-}e^{-4i\left( \alpha _1+\alpha _2 \right)}+o\left( 1 \right).$$\\
$(\text{c})$ When $l_{1,3}$ decays and $l_{2,4}$ is bounded, Eq.~(\ref{s61}) can be rewritten as
\begin{align}\label{s62}
	q_2\left( \chi,\tau \right)=q_{2,-}e^{i\alpha _s}\frac{\det \left( \begin{matrix}
			\left( I-\textbf{1}_{2N}L_{0}^{*} \right) X^*&		-\Gamma L_0\\
			\Gamma ^*L_{0}^{*}&		X^{-1}\\
		\end{matrix} \right)}{\det \left( \begin{matrix}
			I&		-\Gamma L_0\\
			\Gamma ^*L_{0}^{*}&		I\\
		\end{matrix} \right)}, 
\end{align}
where $L_0=\mathrm{diag}\left( 0,l_2,0,l_4 \right)$. Based on the properties of the algebraic cofactor and through appropriate determinant transformations, the solution can be reduced to the expression of one-breather solution corresponding to discrete spectrum $\zeta_2$,
$$q_2\left( \chi,\tau \right) =q_{2}^{1}\left( \chi,\tau \right) +o\left( 1 \right).$$\\
$(\text{d})$ When $l_{1,3}$ grows and $l_{2,4}$ decays, one can multiply the numerator and denominator of Eq.~(\ref{s61}) by the inverse of $\mathrm{diag}\left( l_{1}^{*},1,l_{3}^{*},1,l_1,1,l_3,1 \right)$ from the right. The quotient is expressed as 
\begin{align}\label{63}
	\frac{\det \left( \begin{matrix}
			\left( \tilde{L}_{0}^{*}-\textbf{1}_{2N}L_{0}^{*} \right) X^*&		-\Gamma L_0\\
			\Gamma ^*L_{0}^{*}&		X^{-1}\tilde{L}_0\\
		\end{matrix} \right)}{\det \left( \begin{matrix}
			\tilde{L}_{0}^{*}&		-\Gamma L_0\\
			\Gamma ^*L_{0}^{*}&		\tilde{L}_0\\
		\end{matrix} \right)}, 
\end{align}
where $\tilde{L}_0=\,\,\mathrm{diag}\left( 0,1,0,1 \right),  L_0=\mathrm{diag}\left( 1,0,1,0 \right).$ It can be obtained through direct calculation that $$q_2\left( \chi,\tau \right)=q_{2,-}e^{-4i\alpha _1}+o\left( 1 \right).$$\\
$(\text{e})$ When $l_{1,3}$ grows and $l_{2,4}$ is bounded, one can multiply the numerator and denominator of Eq.~(\ref{s61}) by the inverse of $\mathrm{diag}\left( l_{1}^{*},1,l_{3}^{*},1,l_1,1,l_3,1 \right)$ from the right. The quotient is expressed as 
\begin{align}\label{64}
	\frac{\det \left( \begin{matrix}
			\left( \tilde{L}_{0}^{*}-\textbf{1}_{2N}L_{0}^{*} \right) X^*&		-\Gamma L_0\\
			\Gamma ^*L_{0}^{*}&		X^{-1}\tilde{L}_0\\
		\end{matrix} \right)}{\det \left( \begin{matrix}
			\tilde{L}_{0}^{*}&		-\Gamma L_0\\
			\Gamma ^*L_{0}^{*}&		\tilde{L}_0\\
		\end{matrix} \right)}, 
\end{align}
where $\tilde{L}_0=\,\,\mathrm{diag}\left( 0,1,0,1 \right),  L_0=\mathrm{diag}\left( 1,l_2,1,l_4 \right)$. We can perform multiplication and division transformations on the rows and columns, respectively, with respect to $a_j$, where
\[
\begin{aligned}
	a_1&=\frac{\zeta _{1}^{*}\zeta _{3}^{*}\beta _{12}\beta _{23}}{\zeta _1\zeta _3},& a_2&=\frac{\zeta _{1}^{*}\zeta _{3}^{*}\beta _{14}\beta _{34}}{\zeta _1\zeta _3},\\
	a_3&=\beta _{12}^{*}\beta _{23}^{*},&a_4&=\beta _{14}^{*}\beta _{34}^{*},
\end{aligned} 
\]
and $\beta _{m,n}=\frac{\zeta _m-\zeta _{n}^{*}}{\zeta _{m}^{*}-\zeta _{n}^{*}}$. Introducing the denotion $$\tilde{l}_2=\frac{\zeta _{1}^{*}\zeta _{3}^{*}l_2}{\zeta _1\zeta _3\left( \beta _{12}^{*}\beta _{23}^{*} \right) ^2}, \,\, \tilde{l}_4=\frac{\zeta _{1}^{*}\zeta _{3}^{*}l_4}{\zeta _1\zeta _3\left( \beta _{14}^{*}\beta _{34}^{*} \right) ^2},$$
the solution can be written as
$$q_2\left( \chi,\tau \right)=\tilde{q}_{2}^{1}e^{-4i\alpha _1}+o\left( 1 \right),$$
where $\tilde{q}_{2}^{1}$ has the same form as a one-breather solution $q_{2}^{1}$ excepting $l_{2,4}$ is substituted by $\tilde{l}_{2,4}$. Based on the analysis, one finds that in this limit, the breathers would undergo displacement after colliding, and the asymptotic displacement would be $\left| \Delta \bar{\zeta} \right|=\frac{\mathrm{ln}\left( \bar{\zeta}_0 \right)}{c_{+,n}\cos \alpha _n}$, with $$\bar{\zeta}_0=\frac{d_+-2\cos \left( \alpha _1-\alpha _2 \right) \left( c_{+,1}c_{+,2}-2\cos \left( \alpha _1-\alpha _2 \right) \right)}{d_++2\cos \left( \alpha _1+\alpha _2 \right) \left( c_{+,1}c_{+,2}+2\cos \left( \alpha _1+\alpha _2 \right) \right)},
$$ and $d_+=d_{+,1}+d_{+,2}$. The velocities $V_1=\frac{2\alpha \sin 2\alpha _1\,\,}{\left( d_{+,1}-2\cos 2\alpha _1 \right) c_{+,1}\cos \alpha _1}$ and $V_2=\frac{2\alpha \sin 2\alpha _2}{\left( d_{+,2}-2\cos 2\alpha _2 \right) c_{+,2}\cos \alpha _2}$ corresponding to breathers can be calculated. An example is also shown in Fig.~7(b), whose parameters are chosen appropriately so that the velocity satisfies $V_1 < V_2$ when $\left| \tau \right| \to \infty$. 

\begin{figure}[H]
	\centering
	\begin{subfigure}[b]{0.45\textwidth}
		\centering
		\includegraphics[width=\textwidth]{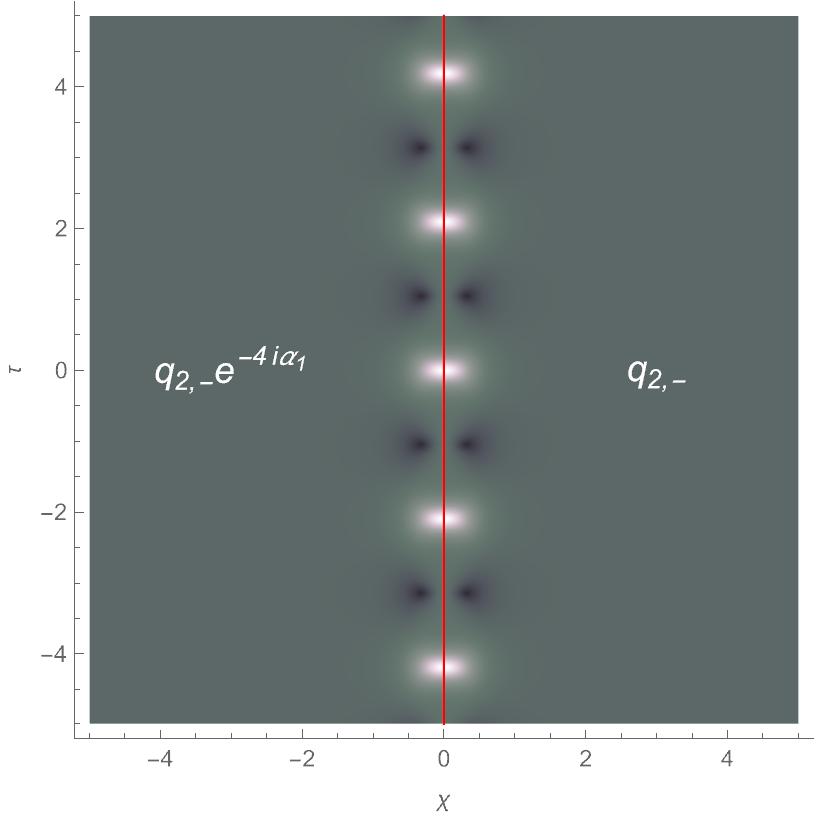}
		\caption{}
	\end{subfigure}
	\hfill
	\begin{subfigure}[b]{0.45\textwidth}
		\centering
		\includegraphics[width=\textwidth]{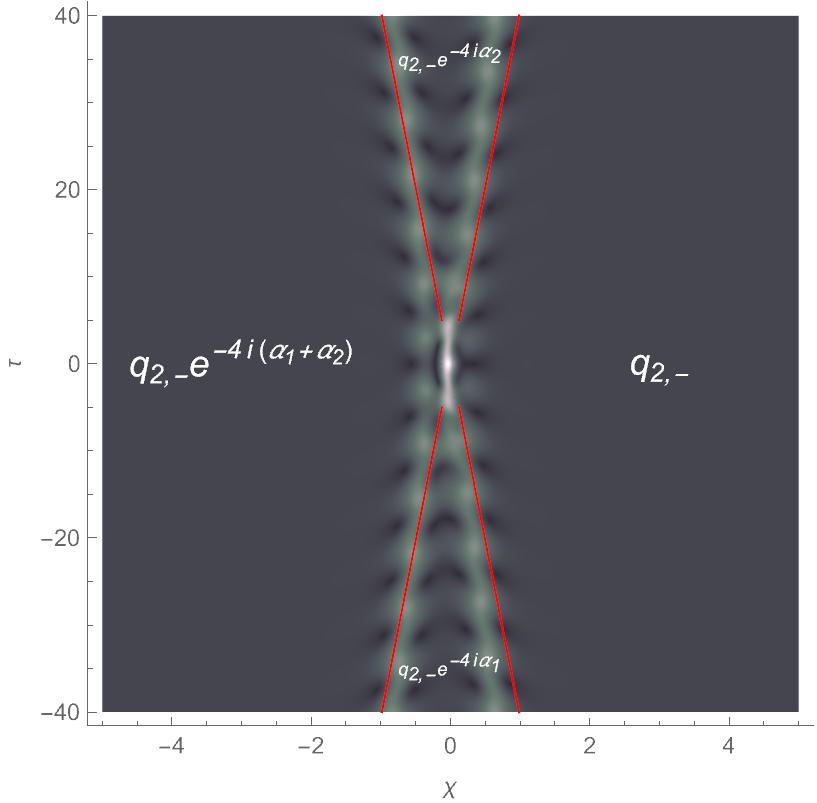}
		\caption{}
	\end{subfigure}
	\caption{(a) Asymptotics of one-soliton solution. $\alpha=0,\ z=2\sqrt{2}$, the red line stands for vecolity. (b) Soliton interactions for two-soliton solutions. $\alpha_1=-\frac{5\pi}{6}, z_1=\frac{1}{2},\alpha_2=-\frac{\pi}{6}, z_2=\frac{1}{2}$, the red lines stand for initial vecolity.}
	\label{f7}
\end{figure}

	\section{Conclusions}
	\hspace{0.7cm}In this paper, we have employed the RH approach to obtain solutions for the SHG equation under mixed boundary conditions and have also conducted one-breather solution asymptotics analysis and two-breather solutions  interaction analysis corresponding to the second harmonic solutions $q_2$ in the single-pole case. These studies are elaborated in the following three aspects:
	
	Firstly, we have reconstructed the corresponding scattering eigenfunctions and the scattering potentials with the RH problem. Distinct from the GLM equations \cite{ref20}, we obtain the matrix form of soliton solutions, breather solutions, and soliton-breather solutions, which correspond to both fundamental and second harmonic solutions.
	
	Secondly, the corresponding soliton images illustrate two phenomena dictated by the parameters: solutions manifest as soliton or breather types, while their interactions range from elastic collisions to the bound states that form as their velocities are nearly identical.
	
	Thirdly, based on the obtained breather and two-breather solutions, we have investigated the interactions between breathers through asymptotic analysis. Given the dynamical properties of breathers, the asymptotic states are categorized into five classes for discussion. Compared with the nonlinear Schrödinger equation \cite{ref37}, one has established that if the velocity directions are opposite, then the resulting displacement directions are also opposite. Moreover, the displacement has been determined analytically.
	
	This work has presented several interesting results, yet several issues remain to be further discussed. In subsequent research, we will endeavor to address the matrix form of the solution in the double-pole case and to rigorously analyze the corresponding soliton-breather interactions in this scenario. Furthermore, the long-time asymptotic behavior of the fundamental and second harmonic solutions remains an important but not fully understood problem. In future work, we plan to apply rigorous analytical methods, such as the Deift-Zhou nonlinear steepest descent method, to conduct a systematic analysis for the asymptotic behaviors of solutions.
	
	\vspace{5mm}\noindent\textbf{Data availability}
	\\\hspace*{\parindent}
	
	No data was used for the research described in the article.
	
	\vspace{5mm}\noindent\textbf{Acknowledgements}
	\\\hspace*{\parindent}We express our sincere thanks to each
	member of our discussion group for their suggestions. This work has been supported by the Fund Program for the Scientific Activities of Selected Returned Overseas Scholars in Shanxi Province under Grant No.20220008, and the Shanxi Province Science Foundation under Grant No.202303021221031.

\end{document}